\documentclass[lettersize,journal]{IEEEtran}
\usepackage{cite}
\usepackage{amsmath,amssymb,amsfonts}
\usepackage{caption}
\usepackage{algorithmic}
\usepackage{graphicx}
\usepackage{textcomp}
\usepackage{longtable}
\usepackage{array}
\usepackage{etoolbox}
\usepackage[normalem]{ulem}
\usepackage{stfloats}
\usepackage{url}
\usepackage{verbatim}
\usepackage[caption=false,font=normalsize,labelfont=sf,textfont=sf]{subfig}
\usepackage{xcolor}
\def\thevol{1}
\def\myyear{2023}
\makeatletter
\patchcmd{\@evenfoot}{2016}{\myyear}{}{}
\patchcmd{\@oddfoot}{2016}{\myyear}{}{}
\makeatother
\usepackage{academicons}

\begin{document}

\title{Physics-Informed Machine Learning for Data Anomaly Detection, Classification, Localization, and Mitigation: A Review, Challenges, and Path Forward}
\author{{Mehdi Jabbari Zideh, \IEEEmembership{Student Member, IEEE},
Paroma Chatterjee, \IEEEmembership{Member, IEEE},  and Anurag K. Srivastava, \IEEEmembership{Fellow, IEEE}}
\thanks{Authors are with the Lane Department of Computer Science and Electrical Engineering, West Virginia University, Morgantown, WV 26505 USA (e-mail: mj00021@mix.wvu.edu, pc00026@mix.wvu.edu, anurag.srivastava@mail.wvu.edu)}
\thanks{This work was supported in part by the U.S. National Science Foundation FW-HTF award 1840192. We would like to acknowledge Dr. Sarika Khushalani Solanki for technical support.}
\thanks{Corresponding author: Mehdi Jabbari Zideh (e-mail: mj00021@mix.wvu.edu).}}


\maketitle

\begin{abstract}
{Advancements in digital automation for smart grids have led to the installation of measurement devices like phasor measurement units (PMUs), micro-PMUs ($\mu$-PMUs), and smart meters. However, a large amount of data collected by these devices brings several challenges as control room operators need to use this data with models to make confident decisions for reliable and resilient operation of the cyber-power systems. Machine-learning (ML) based tools can provide a reliable interpretation of the deluge of data obtained from the field. For the decision-makers to ensure reliable network operation under all operating conditions, these tools need to identify solutions that are feasible and satisfy the system constraints, while being efficient, trustworthy, and interpretable. This resulted in the increasing popularity of physics-informed machine learning (PIML) approaches, as these methods overcome challenges that model-based or data-driven ML methods face in silos. This work aims at the following: a) review existing strategies and techniques for incorporating underlying physical principles of the power grid into different types of ML approaches (supervised/semi-supervised learning, unsupervised learning, and reinforcement learning (RL)); b) explore the existing works on PIML methods for anomaly detection, classification, localization, and mitigation in power transmission and distribution systems, c) discuss improvements in existing methods through consideration of potential challenges while also addressing the limitations to make them suitable for real-world applications.} 

\end{abstract}




\section{Introduction} \label{sec:introduction}

\IEEEPARstart{R}{eal}-time monitoring and control in electric grid is critical for the reliable and efficient operation. Advancements in measurement devices, such as phasor measurement units (PMUs), provide high-resolution measurements for decision support, however, handling the significant amount of data generated by these devices poses a challenge for decision-making processes. Although it enhances the understanding of the system's dynamics by offering better visibility, there is also an increased probability of encountering abnormal measurements, such as outliers or anomalies. The collected data must be processed for control carefully to prevent any unnecessary and untimely action.
The detection and classification of anomalous data help enhance situational awareness, allowing system operators to take appropriate actions.

The complexity of physics-based methods makes them unsuitable for modeling the behavior of very large interconnected systems, especially when dealing with a significant amount of data. It is worth mentioning that in many real-world applications, a descriptive and universally acceptable physical model of the system is either not available, or difficult to acquire, leading to simplification of the dynamics \cite{b3}. Moreover, with the increasing availability of data, processing a massive amount of data would be computationally expensive \cite{b4}. Hence, purely data-driven methods cannot be a solution for power system operation and management. Furthermore, machine learning (ML) methods, which provide high-resolution solutions by capturing spatiotemporal correlations and complex patterns in available training data \cite{b5,b6,b7,b8,b9}, can produce infeasible and unsatisfactory outputs not satisfying physical system constraints, especially when dealing with dynamic systems. The main reasons for their limited success are \cite{b10,b11}: (i) inability to generalize to out-of-sample datasets by only capturing the intrinsic features of available training or input data, (ii) inability to learn system limitations solely based on available data, thereby producing physically infeasible or scientifically inconsistent results, (iii) lack of explainability and interpretability of the results {owing to the black-box nature of ML methods}, and (iv) need for labeled data, which is usually difficult to obtain for real-world applications.

These limitations motivated {PIML} methods that integrate system physics with ML, to leverage strengths of physics-driven {interpretability} with ML-driven faster computational time (once trained) and scalability. The PIML methods are expected to capture the dynamic nature of physical systems {to} help ensure the solutions' feasibility {and} explore inherent temporal and spatial features in their operation \cite{b10,b12,b13,b14,b15,b16}. 

{In this paper, existing strategies to embed physical knowledge in ML methods are reviewed, followed by the applications of the existing PIML methods in power systems, specifically a relatively exhaustive survey of studies that employed the PIML approaches for anomaly detection, classification, localization, and mitigation in power transmission and distribution systems. The key difference between this review paper and previous survey papers on PIML methods is the application domain of application (see Section \ref{piml_surveys} for more details). To the best of our knowledge, this work is the first literature review paper on the applications of PIML methods in power systems with the main emphasis on anomaly detection, classification, localization, and mitigation. This paper provides a thorough understanding of current progress, potential challenges, and possible solutions for the reliable and secure operation of cyber-physical systems.}

The main contributions of this work {are summarized as:}
\begin{itemize}
\item{First, a comprehensive review of different strategies and techniques used to incorporate the physics of the system into ML approaches (supervised, semi-supervised, unsupervised, and RL) is provided.}
\item{{Second, previous literature review papers of PIML methods and their applications in different scientific fields are provided}}
\item{Third, a survey of PIML methods for anomaly detection, classification, localization, and mitigation in power transmission and distribution systems is discussed.}
\item{Finally, the potential challenges, {possible solutions,} and the path forward are discussed to improve {the} applications {of PIML methods} in real-world systems.}
\end{itemize}

The remainder of this review paper is organized as follows. Section \ref{pimls} introduces different types of PIML methods and provides the details of how to incorporate physical laws into each step of these approaches. {Previous review papers of PIML methods in different scientific fields are presented and analyzed in Section \ref{piml_surveys}.} Section \ref{piml_adcl} reviews the current research works that use PIML for anomaly detection, classification, localization, and mitigation in power transmission and distribution systems. {Potential challenges including} {t}he spatiotemporal feature extraction in input data using the proposed PIML methods {, possible solutions,} and a path forward for applications of these methods in real-world cases are provided in Section \ref{disc}. Finally, the summary is provided in Section \ref{concl}.

\section{Physics Informed (PI) Machine Learning Strategies} \label{pimls}


Incorporating the physics of a system into ML algorithms was initially introduced in \cite{b17}, where the authors proposed a NN to find high-quality solutions for a set of partial differential equations (PDEs) for both linear and nonlinear problems.
{Typical machine-learning algorithms are data-driven, using statistics to find patterns in large datasets, to essentially imitate the system or process the generated data. However, that does not always translate into learning the physical limitations of the system. The physical laws governing a system, validated theoretically and empirically over the years, contain the interpretation of natural phenomena along with the abstraction of human behavior in scientific and engineering applications. The idea of introducing this knowledge into existing machine-learning algorithms falls under the broad umbrella of \textit{Physics-Informed Machine Learning}. More specifically, the term "physics-informed machine learning" refers to the utilization of mathematical principles, empirical observations, or physical laws to form prior knowledge and apply them to ML models to enhance their capabilities for finding solutions in complex physical systems. PIML methods are more applicable to real-world scenarios where the physical laws are only partially understood, and there is limited data availability compared to physics-based models, which require a complete understanding of mathematical models, and purely data-driven methods that rely on a vast amount of data. The unique combination of the strengths of both methods results in PIML models that possess interpretability, generalization, and physical consistency. Some of the broader aspects of PIML, as demonstrated by various researchers, have been encapsulated in this review paper.} 

PIMLs were proposed as promising techniques to leverage prior knowledge for finding solutions \cite{b18} and a class of data-driven solvers for solving nonlinear PDEs \cite{b19}. In general, four techniques have been proposed to integrate physical laws into ML models to make them physically meaningful \cite{b10}, \cite{b13}, \cite{b20}, \cite{b21}: (i) physics-informed loss function, (ii) physics-informed initialization, (iii) physics-informed design of architecture, and (iv) hybrid ML-physics models.
This section provides details of incorporating physics into different stages of supervised/semi-supervised, unsupervised, and RL methods.

\subsection{PI Strategies in Supervised and Semi-supervised Learning Methods} \label{piml_s_ss}

Supervised learning approaches use input-output pairs for constructing a model to map the inputs to their desired outputs \cite{c28}. Data samples are separated into training and testing samples to develop the model and evaluate its performance for prediction or classification. Supervised learning methods use labeled data to train and test a model but in cases where labeled data is scarce or access to the labeled data is difficult due to security reasons, semi-supervised methods utilize a combination of labeled and unlabeled data to make better predictions or classification \cite{a29}. That is, they use additional relevant information provided by the unlabelled data to increase the performance of the learning algorithms. Considering both supervised and semi-supervised methods in one category, the physical meanings could be applied by the same strategies to both methods.

Due to the lack of sufficient real-world data, generating synthetic data using generative models or physics-based methods can help prepare input data for training and evaluating ML models, as can be seen in \cite{b28} {and} \cite{b36}. However, for the purpose of this paper, the focus is on models where physics is incorporated into the architecture of the ML models.

\begin{figure*}
    \hspace{0mm}
    \centering
    \captionsetup{justification=centering}
    \includegraphics[clip,trim=1.1cm 0.7cm 3.2cm 1cm, width=0.99\textwidth]{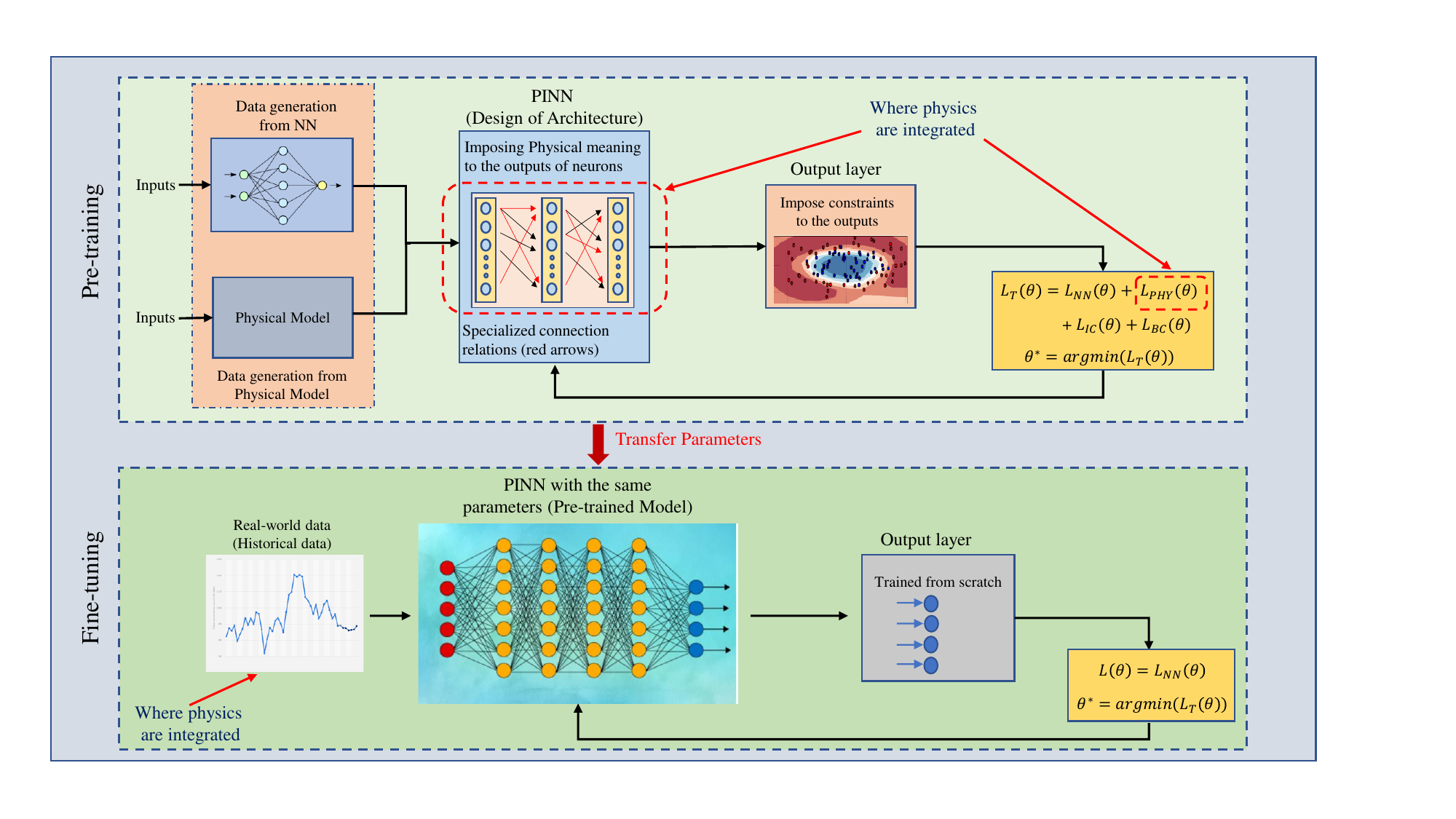}
    \caption{\small Physics-informed strategies for supervised/semi-supervised learning methods.}
    \label{fig:supervised}
\end{figure*}

\emph{\textcolor{blue}{\textbf{ 1) Physics-Informed Architecture:}}}
The integration of physical concepts into the input data does not alleviate the black-box nature of ML methods. In cases where generating realistic synthetic data is not possible, embedding the meaningfulness to the design of the internal architecture of ML models increases their interpretability. As shown in Fig. \ref{fig:supervised}, this can be imposed on the outputs of neurons or by making specialized relations in the connections. As an example, in \cite{b26}, authors developed a framework to impose the prior knowledge on the structure of NNs and also enforced physical constraints on the internal states and outputs of their model to learn the modeling of dynamical systems. Another example of the physics-informed design of architecture is \cite{b27}, where Daw $et$ $al.$ introduced physical meanings to the connections of NNs to ensure physical consistency and apprehend physics-based relationships in the network for lake temperature modeling.

\emph{\textcolor{blue}{\textbf{2) Physics-Informed Loss Function:}}}
Another way of incorporating physical concepts into ML algorithms is to impose constraints on the outputs of these methods and form a loss function representing the deviations of predicted outputs from the physical laws. In general, the dynamics of the system are represented through a set of PDEs or algebraic equations which are approximated as a PINN as outlined in \cite{b19} and \cite{c19};
\begin{equation}\label{eq1}
f{\left(t, x  \right)}=\mathcal{N\left [ \mathit{u\left ( t,x \right )} \right ]}, x\in \textit{X}, t\in \left [ 0, T \right ]
\end{equation}
where $x$ is the input variable, $t$ is the time variable, $\mathcal{N}$ creates physics-informed differential or algebraic equations describing the physical principles of the system, $u$ is an unknown function approximated by a NN, and $f(.)$ is a PINN. As shown in Fig. 1, the total loss including data loss, initial and boundary conditions loss, and violations from governing physics-informed equations is defined as
\begin{equation}\label{eq2}
\mathcal{L_{T}\left( \theta \right)}=\mathcal{L_{NN}\left( \theta \right)} + \mathcal{L_{PHY}\left( \theta \right)}+\mathcal{L_{IC}\left( \theta \right)}+ \mathcal{L_{BC}\left( \theta \right)}
\end{equation}
where $\theta$ represents the parameters. The first term representing the loss originated from the deviations of the predicted output $u_{NN,pred}$ from the labeled output $u_{true}$ in a supervised manner is given by
\begin{equation}\label{eq3}
\mathcal{L_{NN}}\left( \theta \right) =\frac{1}{N_{n}}\sum_{k=1}^{N_{n}}\left| u_{NN,pred}^{k}-u_{true}^{k} \right|^{2} +\lambda_{NN}R\left( \theta \right)
\end{equation}
where $N_{n}$ is the data size, $\lambda_{NN}$ is a hyper-parameter preventing the weights from overgrowing, and $R$ is regularization. The collocation points loss, boundary and initial points losses are expressed as
\begin{equation}\label{eq4}
\mathcal{L_{PHY}\left( \theta \right)} =\frac{1}{N_{f}}\sum_{k=1}^{N_{f}}\left|\mathcal{N} \left [u{( t_{f}^{k},x_{f}^{k}} \right )]-f( t_{f}^{k},x_{f}^{k}) \right|^{2}
\end{equation}

\begin{equation}\label{eq5}
\mathcal{L_{IC}\left( \theta \right)} =\frac{1}{N_{IC}}\sum_{k=1}^{N_{IC}}\left|u \left ( 0,x_{IC}^{k} \right )-g( x_{IC}^{k}) \right|^{2}
\end{equation}

\begin{equation}\label{eq6}
\mathcal{L_{BC}\left( \theta \right)} =\frac{1}{N_{BC}}\sum_{k=1}^{N_{BC}}\left|u \left ( t_{BC}^{k}, x_{BC}^{k} \right )-h( t_{BC}^{k},x_{BC}^{k}) \right|^{2}
\end{equation}
Here $N_{f}$, $N_{IC}$, and $N_{BC}$ are sets of collocation points, initial conditions values, and boundary conditions values, respectively. $g(.)$ and $h(.)$ denote initial and boundary conditions, respectively. Compared to the traditional loss function that uses labeled data for computing the deviations, the physical term does not need labels and works for both supervised and unsupervised methods. The integration of physics-based terms into loss functions has shown benefits as shown in \cite{b22} and \cite{b23}. In \cite{b22}, a physics-based loss function is constructed and added to the main loss function by using the physical meanings between density, temperature, and depth of water to model the water temperature in a lake. Authors in \cite{b23}, considered the deviations from partial differential equations as a penalty term to reflect the physical meanings in the loss function. Both papers denoted the advantages of the applied PIML methods for better generalization, scientific consistency, and improved accuracy of the results.

The last technique applied to ML methods to be physically feasible known as physics-informed initialization leverages the benefits of transfer learning. In this strategy, one of the aforementioned PIMLs is applied to pre-train the model, and then in the fine-tuning stage, all the parameters (weights and biases) are transferred from the pre-trained model while the output layer is trained from scratch with limited real-world data. As shown in Fig. \ref{fig:supervised}, in the fine-tuning stage, since real-world data is based on the physics of the system, the loss function does not have the physics-based loss term and only data loss is considered. This strategy was successfully employed in \cite{b25} where Jia $et$ $al.$ used physics-informed initialization to first train a PINN by simulated data from the physical model and then fine-tune the final model using limited real observed data for the lake temperature prediction.

These strategies may be employed separately as mentioned in each section or a combination {of these methods} can be used for better generalization to observe unseen scenarios as shown in \cite{c25} and \cite{d25}. Authors in \cite{c25} proposed a doubly fed cross-residual network based on magnetic flux leakage (MFL) defect quantification theory and also applied this theory in the NN loss function to train a model for quantifying MFL defects. Sun $et$ $al.$ \cite{d25} proposed a PINN design of architecture using the original inputs and their corresponding physical outputs, and a PI loss function based on the concepts of the ultrasonic guided wave to detect microcracks and quantify their depth, length, and direction.

\subsection{PI Strategies in Unsupervised Learning Methods} \label{piml_us}

As opposed to  supervised learning methods, unsupervised learning approaches aim at capturing the explanatory patterns and features in raw data samples without the need for any labels \cite{e25},\cite{f25}. 
In this section, strategies for incorporating physics into two types of most commonly used unsupervised learning approaches i.e., generative models and clustering, are presented.

\emph{\textcolor{blue}{\textbf{1) Physics-Informed Generative Models:}}}
Generative modeling is one of the common types of learning approaches to unsupervised learning. In these approaches, the goal is to learn a model to find a probability distribution of input data samples $(p_{model})$ to be as similar as possible to the original data distribution ($p_{data}$) \cite{g25}. Two common generative models are variational autoencoders (VAEs) and generative adversarial networks (GANs). For integrating physical concepts into these kinds of unsupervised methods, the strategies for supervised learning explained in Section \ref{piml_s_ss} can be employed. Here, the general idea of creating the model is similar to the physics-informed supervised/semi-supervised learning methods, for example, to incorporate physics into the input data, an unsupervised method for data generation can be used, or the design of architecture could be changed while using an unsupervised algorithm. However, for integrating physical meanings into the PINN loss function, some changes are proposed in the following.

Since only unlabeled data is available, the supervised loss function changes to unsupervised loss minimization where two terms are involved as follows:
\begin{equation}\label{eqs7}
\mathcal{L}_{g}\left( \theta \right) =\frac{1}{N_{g}}\sum_{k=1}^{N_{g}}\left| x_{i}^{k}-x_{rec}^{k}(\theta)  \right|^{2} +\lambda_{g}R\left( \theta \right)
\end{equation}
\begin{equation}\label{eqs8}
\mathcal{L}_{f}{\left( \theta \right)} =\frac{1}{N_{f}}\sum_{k=1}^{N_{f}}\left|\mathcal{N} \left (t_{f}^{k}, x_{rec}^{k}(\theta) \right)-f( t_{f}^{k},x_{rec}^{k}(\theta)) \right|^{2}
\end{equation}
where Eq. \ref{eqs7} and Eq. \ref{eqs8} represent the reconstructed (generated) data loss and physics-informed loss functions, respectively. $x_{i}$ is the input data, $x_{rec}(\theta)$ is the reconstructed output by the model, $N_{g}$ is number of data points, and $N_f$ is number of collocation points. Other variables and parameters are the same as what were mentioned in Section \ref{piml_s_ss}.

\begin{figure}[ht]
    \centering
      \includegraphics[clip,trim=6cm 2.5cm 6cm 1cm, width=1.25\linewidth]{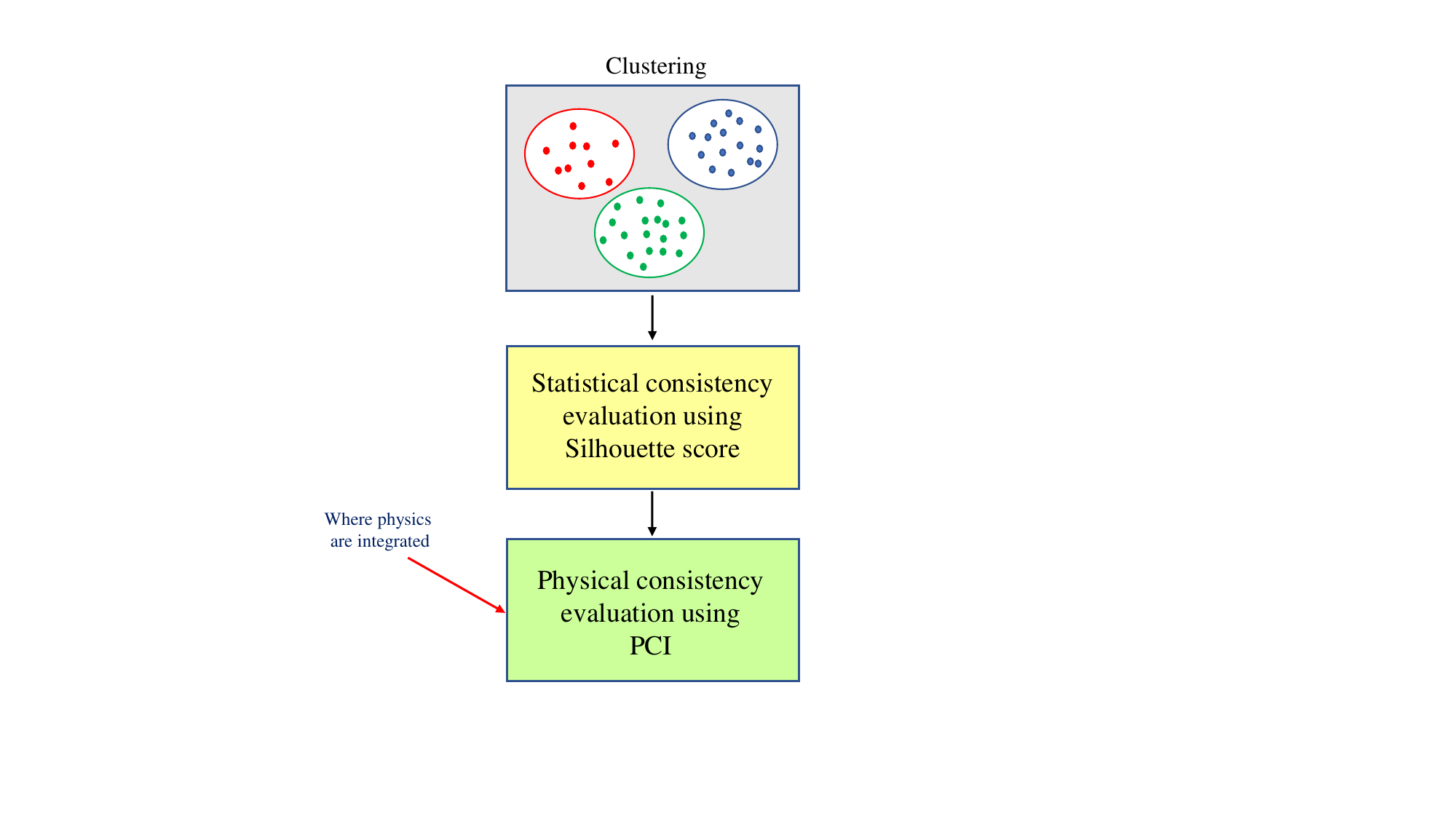}
 \caption{\small  Diagram of the physics-informed clustering method.}
 \label{fig:PI_Clustering}
\end{figure}

\emph{\textcolor{blue}{\textbf{2) Physics-Informed Clustering:}}} Clustering is a class of unsupervised learning methods used for grouping raw data samples with the same statistical characteristics. The aim is to propose a general framework to make these strategies both statistically consistent and physically meaningful. Fig. \ref{fig:PI_Clustering} shows the workflow that can be used to make physics-informed clustering methods. For this purpose, first, input data samples are grouped using one {of} the clustering methods. Then, to assess the quality of a cluster from the statistical point of view, one of the quality metrics mentioned in \cite{h25} can be used, and ultimately, to make the data points physically consistent, their feasibility is evaluated using the underlying physical rules. The  quality assessment metric used here for the statistical consistency is the Silhouette coefficient strategy \cite{i25}, which provides the optimal number of clusters based on the similarity of each data point to other points in its own cluster and dissimilarity to other clusters. This score for data point $x$ in $j$-th cluster is defined as:
\begin{equation}\label{Sil}
S_{s}^{x}=\frac{b_{x}-a_{x}}{max(a_{x},b_{x})}, x\in X^{c_{j}}  
\end{equation}
where $c_{j}$ is the $j$th cluster, $X^{c_{j}}$ is the set of points in the $j$th cluster,  $a_{x}$ (similarity index) and $b_{x}$ (dissimilarity index) are the average distances of data point $x$ from all the data points in the same cluster and the next closest cluster, respectively.

Silhouette score ranges from -1 to 1; the closer to 1, the higher chance for the data point to be well-clustered. The negative values of this score mean that those data points are not correctly assigned to their current cluster. Therefore, this score is used for improved clustering, i.e., the data points with negative Silhouette scores are moved to their nearest neighboring cluster.
To make the data points in each cluster physically consistent, the dynamics of the system are represented by a set of algebraic or differential equations defined in Section \ref{piml_s_ss}.

After evaluating the statistical consistency, the data points need to satisfy the underlying equations of the system. Therefore, a physical consistency index (PCI) for data point $x$ in $j$-th cluster is defined for the physics-informed clustering method:
\begin{equation}\label{clustering}
S_{p}^{x}=|\mathcal{N}{\left (t,x)  \right|} ,\ x \in X^{c_{j}} 
\end{equation}
where higher values of PCI correspond to larger deviations from the physics of the system whereas the physically consistent data points have zero value PCI. The physics-informed clustering strategy was employed in \cite{j25}, where the authors proposed the Silhouette score to evaluate the statistical consistency of data points in each cluster and also a new physics-based parameter considering the concepts of displacement-discontinuity theory {was introduced} for visualization of geo-mechanical alteration distribution to detect spatial variations of shear waves.

\subsection{PI Strategies in Reinforcement Learning Methods} \label{piml_rl}

Reinforcement learning has emerged as an efficient tool for solving complex decision-making processes in various fields such as self-driving cars, cyber-physical systems, biomedical applications, finance, computer vision, speech recognition, and game theory \cite{suttonbartorl,lideeprl,MnihKSGAWR13}. Traditionally, RL algorithms receive states of a system as input and learn the environment by exploring an unconstrained action space to determine the optimal policy that maximizes rewards \cite{suttonbartorl}.  

RL can be defined by a mathematical framework known as the Markov decision process (MDP), which provides a way of modeling reward and state transition resulting from a particular action performed by an agent. An MDP comprises the following fundamental components: a set of discrete states ($\mathcal{S}$), a set of actions an agent can make ($\mathcal{A}$), a set of reward signals (usually real numbers, $\mathcal{R}$), and finally a set of transition probabilities. RL methods are, in essence, solutions to finding the optimal policy for MDPs. 

A desirable model {chosen for an environment} will be one closely mimicking the real world {and} can be perceived as a set of finite numbers of discrete states, $\{s_t\} \in \mathcal{S}$. A single agent interacting with the world chooses to perform an action $a_t \in \mathcal{A}$ {to influence} the environment, in order to yield maximum rewards. For a given set of discrete states and actions, the number of policies that can be executed are $|\mathcal{A}|^{|\mathcal{S}|}$, where a \emph{policy} is a mapping of actions that can be taken given a state, and $|\mathcal{A}|$ denotes the cardinality of set $\mathcal{A}$. The transition from one state to another is governed by a transition probability matrix $\mathcal{P}$ of dimension $|\mathcal{S}|\textrm{x}|\mathcal{S}|$.

In cases of large number of state-action pairs, resulting in large computational costs and memory issues, RL is combined with deep neural networks (DNNs) for learning optimal action-value (Q-value) function or policy. Deep reinforcement learning (DRL) was first introduced by Mnih $et$ $al.$ \cite{x27} by developing a deep Q-network to learn optimal policies on the Attari 2600 games. It then was applied to several applications such as anomaly detection \cite{w27}, cyber security \cite{w28}, digital twin networks \cite{w29}, and energy management \cite{w30,w31}. Although DRL has achieved success, it lacks the comprehension of physical laws, resulting in practically infeasible solutions.

\begin{figure*}[ht]
    \hspace{0mm}
    \centering
    \includegraphics[width=1\textwidth]{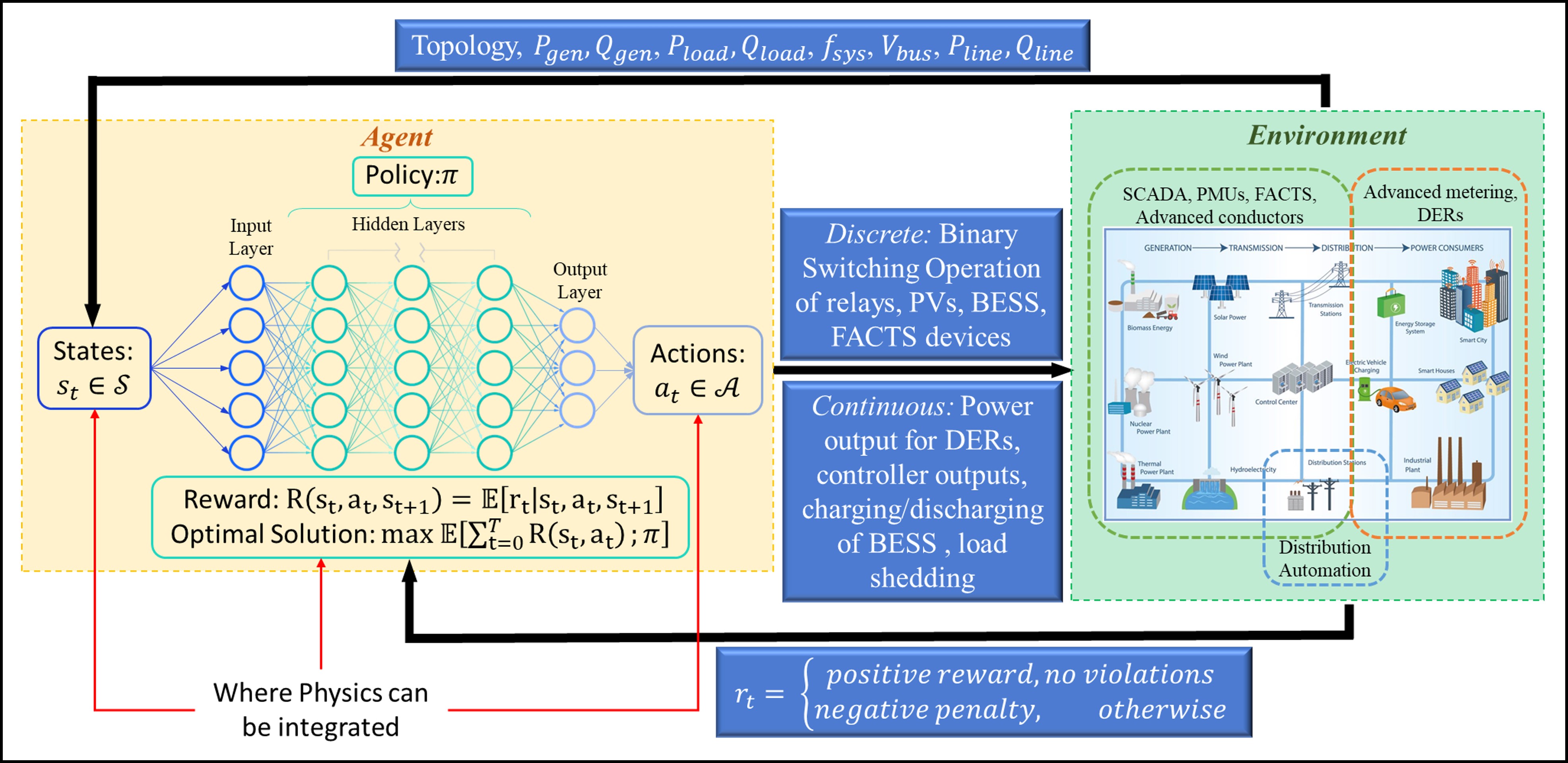}
    \centering
    \caption{\small Physics-informed strategies for Reinforcement Learning methods.}
    \label{fig:PIDRL}
\end{figure*}

In power grids, since maintaining stability is of utmost importance, utilizing system information for optimizing the algorithm ensures an efficient and realistic control action taken by the agent. Physics-informed RL (PIRL) algorithms can be trained faster and used for online applications. Fig. \ref{fig:PIDRL} shows the outline of a typical PIRL model applied to the Smart Grid. This section focuses on explaining how physics-based enhancements can be advantageous for specific applications in the power grid.

\emph{\textcolor{blue}{\textbf{1) Physics-Informed Reward Function:}}} 
The Markov property for state transitions can be assumed to be: $P(s_{t+1}|s_t,a_t\dots s_1,a_1) = P(s_{t+1}|s_t,a_t)$, which states that the probability of transition to the next state depends on the current state and action taken. The reward function is defined as $R(s_t,a_t,s_{t+1}) = \mathbf{E}[r_t|s_t,a_t,s_{t+1}]$, where $\mathbf{E}[.]$ is the expectation of a random variable. Here, $r_t$ is the reward (stochastic or deterministic) at time step $t$, that will be received if action $a_t$ is taken and the state transitions from $s_t$ to $s_{t+1}$. 

Since rewards are given after executing each action, an agent wants to perform actions that consider future rewards. Giving equal weights to all rewards perceived makes it hard for the agent to observe whether immediate or future actions result in maximum returns. A well-known technique to address this weight-assignment issue is to discount the sum of future rewards. Assuming that, at time step $t$, looking forward up to the horizon $T$, the discounted sum of rewards is defined as: $R^{\,\text{disc}} = \sum_{i=t}^{T-1}\gamma^{i-t}r_i,\,\,$, where $\gamma \in[0,1]$ is considered as the discount factor. For $\gamma =1$, the weights of all rewards are considered to be the same, and for $\gamma=0$, it becomes an immediate reward model. The average-reward model can be viewed as a limiting case of the discounted reward model case where $\gamma=1$. It is interesting to see that these reward models can be viewed as different optimality criteria for the agent to pursue a policy that maximizes different forms of function $R^{\text{disc}}$.

To incorporate physical concepts into the algorithm, the deviations of the predicted values from the physically consistent values are computed for each taken action. Therefore, in PIRL, the goal of the agent is not only to maximize its cumulative reward but also to respect the dynamics of the system. As mentioned in Fig. \ref{fig:PIDRL}, the reward function in PIRL can be represented as follows.
\begin{equation}\label{eq7}
R_{t}\left ( a_{t}, s_{t} \right )=\lambda_{phy}r_{phy}\left ( a_{t}, s_{t} \right )+\lambda_{data}r_{data}\left ( a_{t}, s_{t} \right )
\end{equation}
where $\lambda_{phy}$ and $\lambda_{data}$ are predefined parameters based on their importance, $r_{phy}$ and $r_{data}$ represent physics-informed and data-driven reward functions, respectively, defined as 
\begin{equation}\label{eq8}
r_{phy}\left ( a_{t}, s_{t} \right )=-|\mathcal{N}\left ( a_{t}, s_{t} \right )-Q_{\pi }\left ( a_{t}, s_{t} \right )|^{2}
\end{equation}
\begin{equation}\label{eq9}
r_{data}\left ( a_{t}, s_{t} \right )=-|Q_{\exp }\left ( a_{t}, s_{t} \right )-Q_{\pi }\left ( a_{t}, s_{t} \right )|^{2}
\end{equation}
where $\mathcal{N}$ is the same as (\ref{eq1}) and represents the dynamics of the system, $Q_{\pi}(.)$ is the predicted action-value by the policy $\pi$, and $Q_{exp}(.)$ is the experimental Q-value created by experimental studies. Eq. (\ref{eq8}) reflects the deviations of RL outputs from physically meaningful outputs to ensure physical consistency of actions taken, while (\ref{eq9}) shows the differences between predicted outputs and experimental outputs.

As mentioned earlier, the aim of the agent is to maximize its cumulative reward \cite{suttonbartorl}, defined as follows
\begin{equation}\label{eq10}
G_{t}=E[\sum_{k=0}^{ \infty} \gamma^{k}R_{t+k}|s_{t}=s, \pi]
\end{equation}
where $\gamma$ is a discount factor specifying the importance of rewards in a time horizon and $G_{t}$ is the expected cumulative reward earned by the agent following the policy $\pi$.

{Such an idea of incorporating physical principles into the RL method was implemented in \cite{w26} where a PIRL approach was employed to find an optimal policy to estimate the interfacial area concentration (IAC) of the two-phase flows. The authors proposed a physics-informed reward function based on the changes of IAC and the environment of the PIRL was designed through the MDP theory to include the rules of the selected actions and state transformation for capturing the two-phase flow characteristics.}

{A federated multi-agent deep reinforcement learning (F-MADRL) algorithm via the physics-informed reward, for real-time energy management for multiple micro-grids simultaneously, has been developed in \cite{federatedDRL}. The reward is designed as physics-informed to satisfy two targets, realizing the requirements of operation cost and self energy sufficiency of each micro-grid.}

The authors in \cite{GCNN} developed a grid-learning algorithm to control smart inverters for foresighted voltage control. They utilize the power flow equations to develop novel physics-aware Graph Convolutional Neural Network and Graph Recursive Neural Network architectures to capture the spatio-temporal features of the grid signals and deal with the sparsity of data across the network and forecast the states of the system. This information is then used as the states for the DRL algorithm. The control actions for the smart inverters in the network are determined by a stochastic policy that was optimized based on past three-phase voltage phasor information. The reward is determined by the magnitude of voltage deviation from the reference bus. 

For bulk transmission networks, balancing reactive power using non-linear power flow calculations is quite complicated. In Ref.\cite{DRL_Reactive}, a DRL-based reactive power balance method has been proposed. 
For performing demand response, in Ref.\cite{MARLA}, the authors show a simple and flexible way of selecting state elements to reduce the possible number of states based on the physical constraints of a power network, as well as develop a simple equation for the reward function in a reinforcement learning model to reduce the computational complexity of the model.

For adaptive voltage regulation and energy cost minimization, considering uncertainty in renewable generation, loads and energy prices, authors in \cite{SafeDRL} have formulated the optimal operation of a distribution network as a constrained MDP. They define a hybrid action space since the operation of voltage control devices such as capacitor banks, voltage regulators, on-load tap changers, etc, would be discrete (on/off) while the power output of distributed generation and charging and discharging of batteries would be continuous. The safe DRL solution developed here places the nodal voltage constraints, line flow constraints as well as capacity constraints of distributed generation and battery storage units into the policy optimization algorithm.

\emph{\textcolor{blue}{\textbf{2) Physics-Informed Policy:}}} 
Given a correct model of the environment, the next step for an agent is to learn an optimal policy $\pi^*$. Assuming an infinite horizon (number of time steps used to estimate future rewards) discounted reward model ($T\, \rightarrow \, \infty$), the state value function is defined as:
\begin{equation}
\label{eq:value_func}
    \begin{split}
        V_\pi(s) & = \mathbf{E}_\pi\bigg[\sum_{i=t}^{\infty}\gamma^{i-t}r_i\Big|s_t=s\bigg] \\
        & = \sum_a \pi(a|s) \sum_{s^\prime} p(s^\prime|s,a)\bigg[r + \gamma V_\pi (s^\prime)\bigg] \\
        & = R(s) + \gamma \sum_{s^\prime \in S}p(s^\prime|s)V_\pi(s^\prime)
    \end{split}
\end{equation}
where $p(s^\prime|s) = \sum_{a\in A}\pi(a|s)p(s^\prime|s,a)$ is the probability of transitioning from state $s^\prime$ to $s$; $R(s) = \sum_{a\in A}\pi(a|s)R(s,a)$ is the expected reward received when agent starts in state $s$. Here, $R(s,a) = \mathbf{E}[r|s_t=s,a_t=a]$. The probability of taking an action $a$ in state $s$ if the policy is stochastic is given by $\pi(a|s)$. Equation \eqref{eq:value_func} is called the Bellman equation \cite{bellman52} for $V_\pi$. For a known model of the environment, the Bellman equations provide $|S|$ equations for $|S|$ unknown value functions ($V_\pi$). The optimal state value function can be written as
\begin{equation}
    V_*(s) = \max_{\pi }\mathbf{E}_\pi\bigg[\sum_{i=t}^{\infty}\gamma^{i-t}r_i\Big|s_t=s\bigg] = \max_{\pi } V_\pi(s)
\end{equation}
The optimal value function is unique for a given MDP but there can be more than one optimal policy. Similarly, given a policy $\pi$, state-action value function $Q^\pi_t(s,a)$ can be defined for a state $s$ and action $a$ taken at time step $t$, and after following the policy $\pi$, it is defined as
\begin{equation}
    \label{eq:q_pi}
    \begin{split}
        Q_\pi(s,a) & = \mathbf{E}_\pi\bigg[R_t^{\,\text{disc}}\Big|s_t=s, a_t=a\bigg] \\
        & = R(s,a) + \gamma \sum_{s^\prime\in S}p(s^\prime|s,a)V_\pi(s^\prime)
    \end{split}
\end{equation}

The optimal state-action value function is also defined in similar way and is given by $Q_*(s,a) =\max_{\pi}Q_\pi(s,a)$. The optimal state-action value function is unique and will be same for all optimal policies. In practice, most modern RL methods such as Q-learning and SARSA search for policies that maximize the state-action value function. In order to incorporate system information to better guide the optimization problem, network limitations can be incorporated as constraints into the optimization problem along with utilizing experience-based loss functions instead of traditional deep neural networks.

In Ref.\cite{b24}, the authors presented a PINN for RL to model the state formula of generator operation and capture the physical laws of power system by adding physics-based loss term in the loss function representing the deviation from physical laws of the swing equation to train the policy network in RL. This trained model is then employed for choosing appropriate control actions for damping oscillations when disturbances occur in power systems.

During steady state operation of power distribution networks, operational control in smart grid can be modeled as a two-pronged approach: (1) controlling the power output from renewable generation; and (2) switching capacitor banks, on load tap-changing transformer, and other flexible AC Transmission system (FACTS) devices. The authors in \cite{2time} use a two-timescale voltage control approach for fast smart inverter control and slow switching of FACTS devices.

To minimize long-term economic costs for residential smart grid systems, the authors in \cite{RL_Markets} propose a model predictive control (MPC) scheme to approximate a (sub)optimal policy for a multi-agent RL algorithm, that minimizes the economic cost consisting of both the spot-market cost for each consumer and their collective peak power cost, considering local renewable energy production and energy storage.

\emph{\textcolor{blue}{\textbf{3) Physics-Constrained Action Space, Policy \& Reward Function:}}}
In certain applications, it is important not only to constrain the action space based on network information but provide a value function that would be able to explore the viable action space more efficiently or develop a new or constrain an existing reward function to learn the network behavior swiftly and reduce computational complexity. 

In Ref.\cite{RL_PCC}, the authors propose a real and reactive power coordinated control scheme using Prioritized Experience Replay (PER) in a Deep Deterministic Policy Gradient (DDPG) algorithm for online application in distribution networks. They defined the action space based on the real and reactive power outputs for smart inverters and reactive power outputs for SVCs in a given network and placed a large penalty in the reward function for voltage violations.

{A safe multi-agent deep RL based control scheme is proposed in \cite{PS_madrl} for regulating power control of photovoltaics (PVs) by coordinating battery energy storage systems (BESSs) and static var compensators (SVCs) in order to alleviate power congestion and improve voltage quality. They proposed a multi-agent twin delayed deep deterministic (MATD3) policy gradient algorithm with a physics-based shielding mechanism for safe active voltage control in a distribution network with high penetration of PVs.} 

{An actor-critic-based agent with physics-based action-space reduction, state selection, and reward design has been developed in \cite{cbrl}, that controls the power grid’s topology to prevent thermal cascading. They reduced the action space and state space dimensions by analyzing the grid topology and operating conditions and designed a reward function to provide gradients to improve the RL agent when there are overloads in the network.}


\section{Review of Surveys on PIML methods} \label{piml_surveys}

{This section summarizes previous surveys on PIML methods across various scientific domains and differentiates their work from the contributions of this review paper.}

{Authors in \cite{b12} review hybrid model-based ML approaches and their applications in cyber-physical systems (CPS). The metrics for evaluation of these methods, the strategies of fusion of physics-based and data-driven methods, possible challenges, and directions for future work in the CPS domain are comprehensively presented and discussed. The survey in \cite{b13} presents a systematic overview of PINNs in the domain of power systems, focusing on physics-informed loss function, physics-informed initialization, physics-informed design of architecture, and hybrid physics-DL models for state/parameter estimation, dynamic analysis, power flow calculation, optimal power flow, anomaly detection and location, model and data synthesis, and the likes.}

{A literature review of the research studies of PINN methods that utilized different network architectures, the strategies to integrate physical knowledge, and the type of physical information applied by the previous researchers, as well as the applications of these methods in different scientific fields, are given in \cite{b14}. In \cite{big_data}, the authors comprehensively reviewed the most current research studies and the newly proposed techniques with a focus on enhancing PINN methodologies. They categorized these strategies as minimized loss, extended, and hybrid to find how to improve the performance of PINNs. The learning paradigm of PIML, specifically in NN and RL, is systematically reviewed from three perspectives of machine learning tasks, representation of physical prior, and methods for incorporating physical prior in \cite{review_overall} for a broad range of applications in computer vision and robotics control. An extensive study on PIML has been provided in \cite{review_general}, which summarized PIML from the aspects of motivation, physical knowledge, and methods of integration of said physical knowledge in ML. They discussed physics-informed data enhancement, NN architecture design and optimization.}

{Authors in \cite{b20} review the strategies to embed physical principles into ML algorithms. They explain the mathematical representation of PDEs in NNs and how PIMLs can tackle the issues of the high dimensionality of data and uncertainty quantification. The capabilities of these methods in solving problems that are impossible or difficult to solve by the traditional methods are shown using various examples. Researchers in \cite{PDE_survey} have given a survey of the strategies for solving PDEs using DNNs with the current progress on the presented methods and provided the actual applications of PIML methods in various scientific research fields including engineering and medical scenarios. As DNNs need large amounts of data, not always accessible for scientific systems or processes, \cite{review_civil} studies how additional information acquired by enforcing physical norms may be used to train such networks. They investigate physics-informed learning that combines (noisy) data with mathematical models, then implement NNs or other kernel-based regression networks. Here, the behavior of complicated physical systems is learned by minimizing the residual of the underlying PDEs by optimizing the parameters of the neural network.}

{A comprehensive review of strategies to integrate physics into NNs is presented in \cite{Pg_Pe}. The authors categorize the strategies into four frameworks including PINNs, physics-guided neural networks (PgNNs), physics-encoded neural networks (PeNNs), and neural operators (NOs), and explain their mechanisms, applications, and limitations along with the lists the recent research papers that employed these strategies.}

{Ref. \cite{fluid_mech} provides the key concepts of embedding domain knowledge in ML methods and gives a review of the applications of PIML models to fluid mechanics. A case study in fluid flow problems by modifying the structure of PINNs is presented and challenges and recommendations for employing PIML methods in this field are discussed. Flow physics-informed learning, that seamlessly integrates data and mathematical models using PINNs has been reviewed in \cite{review_fluid}. They demonstrate the effectiveness of PINNs for inverse problems related to three-dimensional wake flows, supersonic flows, and biomedical flows. Their goal is to find a new approach to simulating realistic fluid
flows, where some data are available from multimodality measurements whereas the boundary conditions or initial conditions may be unknown, as existing computational fluid dynamics (CFD) solvers cannot handle such ill-posed problems.}

{A systematic review of the most recent progress and applications of PIML approaches in subsurface science are presented in \cite{subsurface} to signify their capabilities in providing more robust and trustable outcomes. The main challenges including data availability, domain transferability, and model scalability with their corresponding solutions are also discussed.}

{Ref\cite{weather} explains the main approaches for integrating physics into ML models and reviews previous research and case studies leveraging complex physical principles for weather and climate modeling.}

{A brief review of the methodologies of PIML approaches for reservoir simulations with the main focus on unconventional reservoirs is provided in \cite{reservoir}. The authors discussed the challenges of applications of PIML models for product prediction, new developments, and their advantages.}

{Authors in \cite{reliability_safety} highlight the strategies of integration of physics into ML models and present a review of PIML methods and their applications in reliability and system safety. The factors influencing the selection of PIML types, the potential challenges, and future directions are also discussed.}

{In \cite{review_aero}, the authors analyze the shortcomings of computational fluid dynamics (CFD) and traditional reduced-order models (ROMs) in aerodynamic data modeling and discuss NN and PINN models to solve high-dimensional PDEs.} 

{Authors in \cite{review_pinn} study and assess state-of-the-art PINNs from different researchers’ perspectives and use the PRISMA framework for a systematic literature review for the computational sciences and engineering domain. They categorized newly proposed PINN methods into Extended PINNs, Hybrid PINNs, and Minimized Loss techniques and outlined various potential future research directions based on the limitations of the proposed solutions.}

{As described in this section, prior works have focused on a multitude of problems, from aerodynamics to computer vision to power systems, using PIML methods. This review attempts to conduct a relatively exhaustive survey of PIML methods (both DL-based and RL-based) in power and energy systems, with a special emphasis on anomaly detection, classification, localization, and mitigation, respecting the operational limitations of power networks, and describing the current challenges faced by researchers while ultimately providing potential research directions for real-world applications.}

\section{Anomaly-aware PIML methods} \label{piml_adcl}

In this section, {we provide a comprehensive review of the previous research studies that applied the PIML strategies explained in Section \ref{pimls} for anomaly detection, classification, localization, and mitigation in power transmission and distribution systems.} The overview for Sections \ref{piml_adcl_s_ss}, \ref{piml_adcl_us} and \ref{piml_adcl_rl} have been provided in {Tables \ref{tbl:Supervised_Table} to \ref{table:reinforcement}}, respectively.

\subsection{Supervised/Semi-supervised Learning Methods} \label{piml_adcl_s_ss}

In \cite{b36}, a new methodology is presented that leverages the potential of machine learning algorithms to generate synthetic PMU data for event classification leading to improved accuracy. The authors used GANs and Neural ordinary differential equations (ODEs) to create eventful PMU data which look realistic and respect the physical laws and constraints of power systems. These generated data are then employed to enrich the original historical dataset for achieving better accuracy for event classification implemented by four famous machine learning algorithms. In the training phase, both GANs and Neural ODEs are trained simultaneously using real historical data to capture temporal and spatial correlations of the input data, and then, for data generation, these trained models are combined where the generator of the GAN model is the input of the Neural ODE. To analyze the created data to be physically meaningful, the authors employed Prony analysis and it was observed that the synthetic and real data have similar settling patterns and ranges.

Authors in \cite{b29} proposed a combined strategy of data-driven (ML) method and physics-based bad data detection test (Chi-squared test) to detect malicious data {originating} from cyber-attacks, i.e., false data injection (FDI) attacks, in smart grids.  In the data-driven analysis, the popular Reed-Xaoli (RX) anomaly detection algorithm \cite{b30} is used where first, normal samples are identified from predictions of state estimation (SE) to create an initial mean and covariance matrix for understanding the normal samples’ distribution, and then based on a constant threshold value, the testing phase of RX algorithm is implemented where the new incoming data are compared to the previous points to update the mean and covariance matrix. It flags the point as an anomaly if the distance between the old and new points is above the defined threshold. In the model-based analysis, the non-linear characteristics of the power system are modeled through non-linear equations using the Weighted Least Squares (WLS) approach where a cost function is minimized, and the cost value is compared to the chi-squared value to identify the anomalies. For the final assessment, due to the independent nature of the prediction of the two methods, their predictions are considered independent to have a joint prediction of anomalies in the data sample.

Ref \cite{b32} created a hybrid approach for anomaly detection by fusing the solutions of the data-driven method and physics model-based strategy. In the data-driven method, an Ensemble Correlation-based Detection (CorrDet) is proposed in which instead of considering the measurements of the whole system, local measurements for each bus are evaluated and the corresponding statistical values (means and covariance matrices) are determined to provide more sensitive bad data detection on a specific spatial domain. For improving the estimation of the data-driven method, a model-based framework is employed using the classical WLS method to model the non-linear dynamics of the system through a set of non-linear algebraic equations. The advantage of using the Ensemble CorrDet method compared to the original CorrDet \cite{b33} is that since neighboring buses are spatially close to each other and are more correlated, determining local means and covariance matrices considering only local regions removes unnecessary computations and provides more accurate statistical estimation for detecting anomalies. The results of all the local detectors are combined with the physics-based model to make the final decision for detecting the anomalies.

Authors in \cite{b34} proposed another version of the Ensemble CorrDet method called Ensemble CorrDet with Adaptive Statistics (ECD-AS) to detect anomalies in the case of false data injections in the time-varying power systems. They claim that since both CorrDet \cite{b29} and Ensemble CorrDet \cite{b32} do not capture the dynamic nature of power system loads  and only provide satisfactory solutions under constant load profiles, they cannot successfully address issues in real-time environments. Instead, ECD-AS, which is the extension of work in \cite{b29} and \cite{b32}, employs adaptive statistical values such as adaptive mean, covariance, and anomaly thresholds to account for the dynamic nature of power systems operating points under different daily load profiles. The proposed method learns a series of statistics for each bus of the system with each new dataset and makes decisions for different load profiles to capture the changing nature of power systems. 

Ref \cite{b37} introduces a real-time and scalable detector based on a graph neural network (GNN) for FDI attack detection {where} system topology is represented by graph adjacency matrix of GNN and locations of metered data are modeled through GNN spatially correlated layers. In the proposed model-based data-driven method, possible weak points of the system are identified through a stochastic gradient descent algorithm, and based on a designed unobservable FDI attack, the performance of the proposed GNN detectors is assessed. These detectors are able to warn the system prior to the {power system state estimation} (PSSE), thus helping the proper operation of power systems through the energy management system (EMS).

Ref \cite{b38} proposed a new framework to detect both parameter and FDI attacks. The framework used the difference between real measurements and predicted measurement values by multi-target multivariate regression model (ML method) as inputs to the ECD-AS algorithm proposed in \cite{b34} to find the abnormal measurements for FDI attack detection. If the system is under FDI attack, ML residuals will be large otherwise they are low. They also employed measurement estimation of both the ML method and physics-based SE to find the residual difference between both methods for parameter attack detection. The reason is that FDI is reflected in both ML and physics-based models whereas parameter attack is only reflected in SE models (low ML residuals). 

The authors in \cite{feature} proposed a physics-based ML algorithm that uses pseudo-supervised learning (PSL) leveraging the physics-based features for the detection of single line to ground (SLG) and adversarial cyber attack where first an unsupervised learning method is performed on the normal data samples to learn normal features based on pseudo labels and then using a regressor supervised learner (random forest), the input features are mapped to those pseudo labels. The anomaly detection is performed by combining the two methods by training a model to minimize the difference between the outputs of the methods. For feature extraction, the authors employed a detection metric (DM) proposed in \cite{AutomatedAD} and \cite{OptimalPlacement} along with statistical and raw physics-based features to consider correlations between 3-phase voltage and current phasors.

Work in \cite{GNN_LF} introduced a new method for locating faults in power grids using two separate GNNs to overcome the challenges of sparse observability and having access to the limited amount of labeled data. For the former issue, the authors proposed a topology-aware GNN by constructing an adjacency matrix based on the shortest distance of the nodes. The outputs of this network are compared to the labels for the training of the network. For the latter issue, another GNN was employed using a different adjacency matrix by zeroing out the unrelated correlations of the elements in the graph embedding of the data samples whose nodes are far away from the fault locations predicted in the first GNN. Finally, they calculated the similarity of all pairs of data samples using a distance metric to improve the accuracy of fault locations. In both GNNs, partial labels are utilized for minimizing the loss function.
The authors extended the work in \cite{GNN_LF} with a comprehensive assessment in \cite{PI_Graph} where they used the same strategy (two-stage GNNs) for locating faults in power distribution systems. Once the adjacency matrices for the two GNNs are computed based on the similarity concept, they are employed for embedding physical meanings to the outputs of hidden layers. The main feature of the proposed adjacency matrices in the two papers is that they are constructed by considering only the neighboring nodes, so only the most important measurements are evaluated to locate the faults for the grid nodes. This leads to better fault location accuracy compared to the original adjacency matrix created by the admittance matrix.

{Authors in \cite{SR_PINN} present a sparse regression (SR) model and PINN algorithm to detect load-altering attacks (LAAs) and identify the attacked nodes in the real-time operation of power grids. Both SR and PINN approaches are trained to learn frequency and voltage phase angles as well as physical principles of the systems modeled through non-linear dynamical equations. The proposed method employs mean-square loss and an additional physics-based loss (differential equations) to achieve two objectives; first, to determine a solution for voltage phase angles and frequency, while the second objective is to estimate the unknown parameters that accurately represent the observed measurements and the structure of the differential equations.}

{The work in \cite{GCN_SA} proposes a combined model-based data-driven method using graph convolutional network (GCN) and the spectral analysis (SA) model for diagnosing faults. The model-based SA method utilizes the adjacency matrix to determine the connections or relationships between measurements and generate an association graph based on the similarities. The created graph and all the measurements are subsequently given as inputs to the GCN model to create a hybrid model in which a weight coefficient is assigned to indicate the importance of the prior knowledge and training measurements.}

{In \cite{MansourLakouraj}, a GCN is proposed that utilizes measurements of voltage magnitudes and angles to capture their spatial and temporal correlations for event classification in power distribution systems integrated with distributed energy resources. The correlation of system nodes is represented through an edge weights matrix denoting the higher correlation of the nodes with shorter paths. The elements of this matrix are defined based on a distance-based metric representing the physical distance of two nodes. In the GCN model, the spatial features are modeled by the weighted adjacency matrix whereas a sample matrix including the recorded features for different time slots is used for temporal representation.
The authors then applied the same strategy in \cite{multi-rate_sampling} for event classification, identification of the affected regions, and event locations. The proposed method uses multi-rate PMU measurements to calculate the weighted adjacency matrix for modeling the spatial correlation of the sensors in the distribution system. The GCN model uses adjacency matrix and multi-rate sampling sensor measurements in spectral graph convolution layers to incorporate spatiotemporal correlations of the measurement to eventually determine the type of the events and the affected regions.}

The hybrid ML-physics strategy proposed in \cite{hybrid_ML-physics} combines the concepts of methodologies in \cite{b32, b34} and \cite{b38}. The authors employ the WLS approach to estimate the system's state vector and ECD-AS to process and compare the adaptive statistics of the bus measurements against each other to classify the data samples. The main difference between this research and the previous works is twofold; first, a three-phase dynamic load is added to the system to simulate the real-world load variations; secondly, an adaptive voting system with a variable threshold is employed to consider the overall decision scores of SE and ECD-AS to compute the final fusion score for abnormal data detection.

{Authors in \cite{LSTM-AE_MTD} develop a hybrid PINN model leveraging the benefits of the Long Short-term Memory (LSTM) and Autoencoders (AE) models for reconstructing the temporal voltage measurements and a physics-based moving target defense (MTD) to detect FDI attacks on SE. In the proposed semi-supervised method, LSTM-AE is trained with offline normal voltage data to identify abnormalities. If a positive alarm is detected, an attack detection approach is activated that uses the attack knowledge of the physics-based MTD algorithm to verify the detection result of the LSTM-AE method in the following SE time segments.}

{Researchers in \cite{CPS_GCN} employ a GCN to model and preserve the topology of power system for detecting and localizing the FDI attacks. Using the adjacency matrix for modeling the physical properties of the system, the proposed GCN model applies shared weights and localized filters to extract the features. Similar to other graph-based convolution methods, the power grid buses and transmission lines are represented by vertices and edges, respectively, and along with the adjacency matrix are fed to the proposed GCN method for loss minimization leading to the precise prediction of FDI attacks.}

{The authors in \cite{natural_perturb} model the relationship between current and voltage measurements before and during edge faults as physical constraints and apply these constraints in the loss function of an adversarial training model to distinguish malicious attacks from natural perturbations such as control inputs and variable loads of the power grids. The trained approach which is used for edge fault location identification not only follows the physical laws between current and voltage but also imposes voltage and current measurements to be in a predefine certain range.}

\begin{table*}[h!]
\caption{\small Supervised/Semi-supervised PIML methods for anomaly detection, classification, and localization in power systems}
\centering
 \begin{tabular}{|>{\centering}p{0.8cm}|>{\centering}p{3cm}|>{\centering}p{6cm}|p{6cm}|} 
\hline
 \textbf{Refs.} &  \textbf{Where Physics are Integrated} &  \textbf{Advantages} &   \textbf{Limitations / Issues} \\ [0.5ex] 
\hline

\vfil Ref\cite{b36} & Neural ODE loss function  
 & {\scriptsize \begin{itemize}
 \itemsep0em
        \item High Computational efficiency
        \item Improvement in event classification accuracy
        \item Captures spatial and temporal correlation in PMU data
    \end{itemize}} 
& {\scriptsize \begin{itemize}
 \itemsep0em
        \item Lack of validation in real-world systems
        \item Lack of assessment of the effects of sample window size on the
proposed method
    \end{itemize}} \\ \hline
\vfil Ref\cite{b29} & Ensemble (Hybrid) strategy
(SE + RX algorithm) 
 & {\scriptsize \begin{itemize}
 \itemsep0em
        \item Better accuracy, precision, and recall on combined method compared to state estimator
    \end{itemize}} 
& {\scriptsize \begin{itemize}
 \itemsep0em
        \item The statistical measurements (means and variances) are not
adaptive and do not change with the operating points of the system
proposed method
    \end{itemize}} \\ \hline
\vfil Ref\cite{b32} & Ensemble (Hybrid) strategy
(WLS + Ensemble CorrDet)  
 & {\scriptsize \begin{itemize}
 \itemsep0em
        \item More accurate statistical estimation, computationally cheaper, and more sensitive AD due to local regions’ measurements
        \item Better AD scores compared to each individual method
    \end{itemize}} 
& {\scriptsize \begin{itemize}
 \itemsep0em
        \item Not considering the load changing nature of power systems
    \end{itemize}} \\ \hline
\vfil Ref\cite{b34} & Ensemble (Hybrid) strategy
(Ensemble CorrDet with adaptive statistics)
(combination of \cite{b29} and \cite{b32})  
 & {\scriptsize \begin{itemize}
 \itemsep0em
        \item Captures dynamic nature of power system loads
        \item Captures spatial and temporal correlations in the measured data
        \item Computationally cheaper and sensitive AD
        \item Better performance metrics compared to other ML classification methods
    \end{itemize}} 
& {\scriptsize \begin{itemize}
 \itemsep0em
        \item Linear prediction of the next samples are not investigated
(prediction is only based on estimated mean and covariance of normal samples)
    \end{itemize}} \\ \hline
\vfil Ref\cite{b37} & Design of architecture (GNN)  
 & {\scriptsize \begin{itemize}
 \itemsep0em
        \item Shows the effects of the architectural differences of NNs on AD performance
        \item Better performance in detection metrics than other ML methods
        \item Outperforms other models in large power systems
    \end{itemize}} 
& {\scriptsize \begin{itemize}
 \itemsep0em
        \item Limited to FDI attacks
    \end{itemize}} \\ \hline
\vfil Ref\cite{b38} & Ensemble (Hybrid) strategy (WLS (state estimator) + multi-target multivariate linear regression model)  
 & {\scriptsize \begin{itemize}
 \itemsep0em
        \item Considers both FDI and parameter attacks
        \item Better precision, recall, and F-1 score compared to SE solutions for both single and mixed attacks
    \end{itemize}} 
& {\scriptsize \begin{itemize}
 \itemsep0em
        \item Mixed scenario of FDI and parameter attacks is not evaluated.
    \end{itemize}} \\ \hline
\vfil Ref\cite{feature} & Ensemble (Hybrid) strategy (Physics-based features + pseudo-supervised learning (PSL))  
 & {\scriptsize \begin{itemize}
 \itemsep0em
        \item No need to lower the dimensionality of dataset
        \item Improvement in overall performance (precision and recall) by adding physics-based features
    \end{itemize}} 
& {\scriptsize \begin{itemize}
 \itemsep0em
        \item Training dataset must be anomaly-free, otherwise, it learns
anomalies as normal behavior
        \item One-class classification (single line-to-ground (SLG)) i.e., lack of multi-class classification (e.g., SLG/2LG/LL/3LG)
    \end{itemize}} \\ \hline
\vfil Ref\cite{GNN_LF} & Design of architecture (GNNs)  
 & {\scriptsize \begin{itemize}
 \itemsep0em
        \item High fault location accuracy rate and F-1 score even if in low label rate of dataset
        \item Higher location accuracy rate compared to other NNs methods for
SLG/2LG/LL
        \item Robust to topology changes and node loads variations
    \end{itemize}} 
& {\scriptsize \begin{itemize}
 \itemsep0em
        \item Not showing the performance with load variations.
        \item Not providing explanation of the structure of the proposed GNNs
        \item Misclassification of some unmeasured nodes in low label rate due to consideration of neighboring nodes in adjacency matrices.
    \end{itemize}} \\ \hline
\vfil Ref\cite{PI_Graph} & Design of architecture (GNNs)  
 & {\scriptsize \begin{itemize}
 \itemsep0em
        \item High fault location accuracy rate and F-1 score even if in low label rate of dataset
        \item Higher location accuracy rate compared to other NNs methods for
SLG/2LG/LL
        \item Robust to topology changes and node loads variations
        \item Better explanation of integrating physical knowledge to GNNs compared to Ref \cite{GNN_LF}
    \end{itemize}} 
& {\scriptsize \begin{itemize}
 \itemsep0em
        \item Misclassification of some unmeasured nodes in low label rate due to consideration of neighboring nodes in adjacency matrices.
        \item Requires the optimal placement strategy of micro-PMUs for
higher fault location accuracy.
    \end{itemize}} \\ \hline

\vfil  {Ref\cite{SR_PINN}} & {Loss function}
 & {\scriptsize \begin{itemize}
 \itemsep0em
        \item {No need for offline training}
        \item {Very quick execution for SR method}
        \item {Higher accuracy and precision of SR in slow and fast dynamics}
    \end{itemize}} 
& {\scriptsize \begin{itemize}
 \itemsep0em
        \item {High execution time for PINN method}
        \item {Needs pre-training for PINN to reduce the execution time}
        \item {Low performance of PINN in systems with slow dynamics}
    \end{itemize}} \\ \hline
    
\vfil {Ref\cite{GCN_SA}} & {Design of architecture (GNNs)}  
 & {\scriptsize \begin{itemize}
 \itemsep0em
        \item {Comparatively high accuracy in small training set due to prior knowledge}
        \item {Consistency of the results with the actual situation in large training size}
        \item {Very high accuracy even without weight coefficient of prior knowledge}
    \end{itemize}} 
& {\scriptsize \begin{itemize}
 \itemsep0em
        \item {Not able to distinguish the measurements in under unbalanced measurement distribution and normal condition}
    \end{itemize}} \\ \hline
       
\end{tabular}   
\label{tbl:Supervised_Table}
\end{table*}

\begin{table*}[h!]
\caption{\small {Supervised/Semi-supervised PIML methods for anomaly detection, classification, and localization in power systems (Cont.)}}
\centering
  \begin{tabular}{|>{\centering}p{0.8cm}|>{\centering}p{3cm}|>{\centering}p{6cm}|p{6cm}|} 
\hline
{\textbf{Refs.}} &  {\textbf{Where Physics are Integrated}} &  {\textbf{Advantages}} &   {\textbf{Limitations / Issues}}\\ [0.5ex]
\hline

\vfil {Ref\cite{MansourLakouraj}} & {Design of architecture (GNNs)} 
 & {\scriptsize \begin{itemize}
 \itemsep0em
        \item {Better accuracy, precision, and recall compared to decision tree and linear regression} 
        \item {Robust to the number of PMUs, missing data, and measurement noise ratio}
    \end{itemize}} 
& {\scriptsize \begin{itemize}
 \itemsep0em
        \item {Assessment for limited types of events}
        \item {Lack of evaluation for real-world cases with different network topologies}
    \end{itemize}} \\ \hline
    
    \vfil {Ref\cite{multi-rate_sampling}} & {Design of architecture (GNNs)} 
 & {\scriptsize \begin{itemize}
 \itemsep0em
        \item {Works for multi-rate samples of PMUs}
        \item {Trained and implemented under load and generation variations}
        \item {High event classification and location identification compared to baselines}
        \item {Robust to the number of PMUs, missing data, and measurement noise ratio}
    \end{itemize}} 
& {\scriptsize \begin{itemize}
 \itemsep0em
        \item {Applications of waveform measurement units need to be explored to address the issue of a limited number of recorded samples of PMUs, especially in transient studies}
    \end{itemize}} \\ \hline

\vfil {Ref\cite{hybrid_ML-physics}} & {Ensemble (Hybrid) strategy
(Ensemble CorrDet with adaptive
statistics) (combination of \cite{b29}, \cite{b32}, and \cite{b34})}  
 & {\scriptsize \begin{itemize}
 \itemsep0em
        \item {Considers a more realistic situation by including 3-phase dynamic loads} 
        \item {Higher accuracy, recall, and F1-score compared to SE, ECD-AS, and ECD with a fixed voting system}
    \end{itemize}} 
& {\scriptsize \begin{itemize}
 \itemsep0em
        \item {Needs to be evaluated for various systems with larger sizes}
    \end{itemize}} \\ \hline
\vfil {Ref\cite{LSTM-AE_MTD}} & {Hybrid model (LSTM-AE and physic-based MTD)}  
 & {\scriptsize \begin{itemize}
 \itemsep0em
        \item {Works in real-time operation under different FDI attack scenarios}
        \item {Higher performance metrics such as attack detection and defense hiddenness compared to Max-Rank and Robust MTDs}
        \item {Lowest reactance change and operational cost to detect the attacks}
    \end{itemize}} 
& {\scriptsize \begin{itemize}
 \itemsep0em
        \item {Generator cost is not considered in the optimization problem of the MTD algorithm}
    \end{itemize}} \\ \hline
\vfil {Ref\cite{CPS_GCN}} & {Design of architecture (GNNs)}  
 & {\scriptsize \begin{itemize}
 \itemsep0em
        \item {Implemented under FDI attacks in large power systems}
        \item {Higher detection accuracy compared to conventional NN algorithms}
        \item {Very low detection time and suitable for real-time FDI attack detection}
        \item {Comparatively better accuracy under higher sparsity of attacks}
    \end{itemize}} 
& {\scriptsize \begin{itemize}
 \itemsep0em
        \item {Different types of cyber attacks need to be considered to prove the performance of the method.}
        \item {The proposed method should be also employed and tested in unbalanced power systems.}
    \end{itemize}} \\ \hline  
\vfil {Ref\cite{natural_perturb}} & {Loss function}  
 & {\scriptsize \begin{itemize}
 \itemsep0em
        \item {Robust to natural perturbations such as control input changes and load variation}
        \item {Distinguish natural perturbations from cyber attacks}
        \item {Not very sensitive to changes in hyper-parameters of physics-based loss term}
    \end{itemize}} 
& {\scriptsize \begin{itemize}
 \itemsep0em
        \item {Does not model the dynamics of the system}
        \item {Assessments for other common perturbations caused by serious events such as line trips and generator failure as well as topology changes are required.}
    \end{itemize}} \\ \hline 

\end{tabular}   
\label{tbl:Supervised_Table1}
\end{table*}

\subsection{Unsupervised Learning Methods} \label{piml_adcl_us}

Ref \cite{b31} develops a robust and scalable tool based on a combination of the knowledge of the system provided by an explicit but reduced-order mathematical model (abstract model) and a data-driven approach for dynamic anomaly detection in power systems. In the proposed method, a high-fidelity simulator describing the mathematical model of the system is presented to represent the dynamics of the system, and in the second phase, a filter for detecting both multivariate and univariate cyber attacks is employed to minimize the effects of the difference between the outputs of the high-fidelity simulator and the abstract model. The data-driven model-based approach performs comparatively better than pure data-driven methods with less computational cost.

In \cite{b35}, a physics-guided deep learning (PGDL) is proposed to estimate the states of the system for both single and multiple snapshots for detecting bad data and defending the system against FDI attacks. For single-snapshot, an autoencoder neural network is employed in which the measurements are fed into the encoder of the network, while the decoder is replaced with the physics of the power system, power flow equations, to reconstruct the input measurements. Then, the deviations between actual and reconstructed measurements are considered for training the proposed network. For multi snapshots considering the historical time-series data, the authors presented the time-series PSSE. In this method, instead of an autoencoder, a type of recurrent neural network i.e., LSTM, is used to learn the dynamic nature {of the} states in the time-series data samples and the power flow equations play the role of state estimator to reconstruct the inputs.

Authors in \cite{b39} proposed a physics-informed convolutional autoencoder (PICAE) that works in an unsupervised manner (with no labeled data) to detect high impedance fault (HIF) in the power distribution systems. They formed the elliptical trajectory of voltage against current measurements to model the physical principles of the HIFs. Using PICAE, they reconstructed the normal voltages and employed an elliptical regularization to constrain the voltages and currents to follow the elliptical trajectory.
Ref \cite{b40} used a physics-based graph signal processing method for bad data detection in power distribution systems. The proposed method employs a physics-based graph Laplacian to transform measured voltage signals from the time domain into the frequency domain. Then, the authors reduced the dimensionality of the signals using T-distributed stochastic Neighbor embedding (t-SNE) and employed density-based spatial clustering of applications with noise (DBSCAN) to identify irregular patterns or bad data in the phase-to-phase and phase-to-neutral data measured by smart meters.

In \cite{RobustFault}, a GCN is proposed for fault location analysis where the model is topology aware {which} preserves the spatial correlations of the buses and considers the measurements of micro-PMUs at different locations. The proposed method consists of graph convolution and fully connected layers and uses the weighted adjacency matrix to find the spectral convolution of the measured signals. The performance of the proposed method is compared to three models i.e., fully connected neural network, random forest, and support vector machine, for three types of faults {; SLG, double-line to ground (2LG), and line-to-line (LL)}, and under three conditions (Gaussian noise, data loss of busses, and random data loss for measured data). The results confirm better performance compared to the mentioned ML algorithms and the robustness of the model to data loss errors and measurement noises.

{
An ensemble physics-based Isolation Forest (physics-iForest) algorithm is presented in \cite{physics-iForest} to detect stealthy FDI attacks created on the generated power signals of wind farms. The algorithm utilizes environmental data such as wind speed and air density as well as turbine parameters to calculate the generated wind power and sets a threshold to compute the transmitted power signals. The computed signals along with sensor data go under the feature extraction process to capture the main patterns and trends. To detect anomalous signals, extracted features from the physics-based method and normal historical data are transferred to the iForest algorithm.
}

{The model proposed in \cite{RBM_DGI} utilizes the physical characteristics of wind turbines and leverages the potential of GNNs and physical-statistical feature fusion to develop an anomaly detection framework for wind turbine data. The authors present the detection framework in a multi-phase approach where the physics-based features of operating conditions of wind turbines are extracted and along with the graph adjacency matrix are fed to the Deep Graph InfoMax (DGI) to train a model for learning the normal behavior of wind turbines. In the detection phase, the Restricted Boltzmann Machine (RBM) is employed that uses the updated nodes feature matrix acquired from DGI to identify abnormal states based on an energy-based criterion.}

\begin{table*}[h!]
\caption{\small Unsupervised PIML methods for anomaly detection, classification, and localization in power systems}
\centering
 \begin{tabular}{|>{\centering}p{0.8cm}|>{\centering}p{3cm}|>{\centering}p{6cm}|p{6cm}|} 
\hline
 \textbf{Refs.} &  \textbf{Where Physics are Integrated} &  \textbf{Advantages} &   \textbf{Limitations / Issues} \\ [0.5ex] 
\hline

\vfil Ref\cite{b31} & Ensemble (Hybrid) strategy
(Abstract model-based + High-fidelity simulator)
 & {\scriptsize \begin{itemize}
 \itemsep0em
        \item Successful detection of univariate and multivariate FDI attacks
        \item Minimized model mismatch
        \item Low computational cost in the data training stage
    \end{itemize}} 
& {\scriptsize \begin{itemize}
 \itemsep0em
        \item False alarm generation in cases of much larger covariances in measurements’ noises
        \item The performance of the proposed method is sensitive to the training data size.
    \end{itemize}} \\ \hline
    
\vfil Ref\cite{b35} & Design of architecture 
 & {\scriptsize \begin{itemize}
 \itemsep0em
        \item Stable, robust, and accurate performance against FDI attacks
        \item Considers single-snapshot and multi-snapshot estimations for FID attacks
        \item Better performance in case of FDI attacks compared to WLS(SE)
    \end{itemize}} 
& {\scriptsize \begin{itemize}
 \itemsep0em
        \item Larger  Root Mean Square Error (RMSE) while considering continuous attacks
    \end{itemize}} \\ \hline
    
\vfil Ref\cite{b39} & Loss function  
 & {\scriptsize \begin{itemize}
 \itemsep0em
        \item Distinguishes high impedance faults from other anomalies using the reconstruction error of the proposed method
        \item More robust to noises compared to other unsupervised ML methods by achieving higher F-1 score 
        \item High accuracy in reconstructed normal events
    \end{itemize}} 
& {\scriptsize \begin{itemize}
 \itemsep0em
        \item Lower detection performance with low observability of distribution systems (detection depends on measured ratio)
    \end{itemize}} \\ \hline
    
\vfil Ref\cite{b40} & Clusters 
(using graph signal processing to create a weight matrix for voltage drops)  
 & {\scriptsize \begin{itemize}
 \itemsep0em
        \item High accuracy and F-1 score in finding erroneous data points in both SLG and LL configurations
    \end{itemize}} 
& {\scriptsize \begin{itemize}
 \itemsep0em
        \item Misclassification of some normal and erroneous data points
    \end{itemize}} \\ \hline
    
\vfil Ref\cite{RobustFault} & Design of architecture  
 & {\scriptsize \begin{itemize}
 \itemsep0em
        \item Outperforms other  ML models with higher accuracy
        \item Robust to measurement noises, and data loss error
        \item Increased robustness to the errors and noises with data augmentation
        \item High stability under various network reconfiguration scenarios
        \item Capable of locating high impedance faults accurately
        \item Resolves the low observability issues of data samples
    \end{itemize}} 
& {\scriptsize \begin{itemize}
 \itemsep0em
        \item Lack of application to locate faults in real-world cases (used simulation-only data)
        \item The performance of the model is lower when voltage phasors are ignored
        \item Ignorance of phase angles in data loss error determination
    \end{itemize}} \\ \hline
    
\vfil {Ref\cite{physics-iForest}} & {Ensemble model (Physics-iForest)
}  
 & {\scriptsize \begin{itemize}
 \itemsep0em
        \item {Higher detection accuracy compared to standalone methods}
        \item {More contribution to physical principle compared to statistical and configuration-based feature extraction}
    \end{itemize}} 
& {\scriptsize \begin{itemize}
 \itemsep0em
        \item {Very close false alarm rate to individual ML and physics-based models}
    \end{itemize}} \\ \hline
\vfil {Ref\cite{RBM_DGI}} & {Design of architecture (GNN)}  
 & {\scriptsize \begin{itemize}
 \itemsep0em
        \item {Capable of capturing the nonlinear correlation of SCADA dataset}
        \item {Captures temporal correlation of dataset with the appropriate size of sliding window}
        \item {Better feature extraction by using graph attention network}
    \end{itemize}} 
& {\scriptsize \begin{itemize}
 \itemsep0em
        \item {Depends highly on the sliding window size}
        \item {Fault propagation mechanism and root cause analysis should be investigated.}
    \end{itemize}} \\ \hline
       
\end{tabular}   
\label{tbl:Unsupervised_Table}
\end{table*}


\subsection{Reinforcement Learning Methods} \label{piml_adcl_rl}

The previous learning methods focus on locating an anomaly and/or classifying it. RL methods acknowledge anomalies such as faults, cyber-attacks, and even catastrophic events, and provide control solutions to isolate and stabilize the connected networks. The models developed in the literature to provide control solutions under anomalous conditions have been provided in this section. 

One big aspect to consider while modeling a power system analysis model is the network's vulnerability to FDI attacks. In Ref.\cite{RL-SCOPF}, the authors propose a vulnerability index to assess such a model. The physics-based DRL model utilizes the generator outputs, load inputs, and the network states (voltages and currents at each bus) as the states, and its action space is governed by the physical limitations of the power network model. The adversarial attack designed in this paper exploits the vulnerabilities of both value-based and policy-based model-free DRL algorithms. In order to bypass any bad-data detectors, the manipulated state is designed such that it satisfies physical equality and inequality constraints of power systems, and ultimately influences the DRL agent to shift from its original outputs. The authors provide a measure of vulnerability in the form of expected performance decay evaluated from the pre-attacked and post-attacked expected reward values. 

To protect the power grid, the overlying communication network needs to be protected from malicious entities as well. The authors of Ref.\cite{SA1} provide a solution to this problem by analyzing the data collected from a smart-grid communication network. They utilize this data to perform security situational awareness by first extracting security situational factors, such as including static configuration information, dynamic operating information and flow information in networks, and using neural networks and game theory to create an awareness of network security situation by means of predicting the development trends of the network in next phase based on historical information. Finally, they use game theory and reinforcement learning to realize security situational awareness for smart grid, by assigning the legitimate users and attackers as the players in the game and using two different but relevant types of game models for the strategies of the players in the games.

Possible denial-of-service attacks on communication networks transferring information between controllers and remote sensors can affect power networks severely. For islanded micro-grids, authors in Ref.\cite{marl_dos} developed a secondary control scheme to achieve frequency regulation and simultaneous state-of-charge (SoC) balancing for Battery Energy Storage Systems (BESSs) by using an asynchronous advantage actor-critic (A3C) based multi-agent deep reinforcement learning (MA-DRL) algorithm with centralized off-line learning with shared convolutional neural networks (CNN).
Ref.\cite{ace} presents an intelligent controller for frequency control application in smart grid using a multi-agent RL model, considering network-induced effects, time delay, and change in communication topology. Once an event is detected in the communication network, it triggers a scheme that decides how to update a particular sample data based on the last transmitted information in order for the frequency control to not be impacted.
 
With increasing extreme weather events, ensuring resilience of the grid is becoming increasingly complicated. Although intermittent, distributed renewable generation can be utilized to provide power to critical loads. In Ref.\cite{RL-events}, the authors propose a two-layer framework for a data-driven model predictive control-based proactive scheduling strategy for micro-grids subject to extreme weather events. The proposed model takes into account both the uncertainty of renewable generation as well as random extreme events that can result in the interruption of grid-connection while eliminating anomalous data obtained from smart meters. They integrated an advanced EMS to enhance resilience when disconnected from the grid and consider multiple objectives for their Proactive Scheduling strategy.

To provide an adaptive control strategy for load shedding {and} mitigate fault-induced delayed voltage recovery, authors in Ref.\cite{RL_Control1} augmented the conventional RL model with a physics-informed module, to constrain the action space in order to avoid unnecessary explorations, achieving better sample efficiency and control robustness. They utilize the transient voltage recovery criterion as prior knowledge to create a trainable action-mask for the RL algorithm.

The expanding smart meter infrastructure in the distribution network brings with it an immense amount of data that needs to be processed intelligently to extract valuable information. The authors in \cite{hmi} propose a human-machine RL framework in the smart grid context to formulate an energy management strategy for electric vehicles and thermostatically controlled loads aggregators, to accelerate the decision-making speed and deal with emergent events, such as a sudden drop of photovoltaic (PV) output.

 \begin{table*}[h!]
\caption{\small Reinforcement Learning based PIML Methods for Performing under Anomalous Conditions}
\centering
 \begin{tabular}{|>{\centering}p{0.8cm}|p{7cm}|p{4cm}|p{4cm}|} 
\hline
 \textbf{Refs.} &  \textbf{Where Physics has been Integrated} &  \textbf{Advantages} &   \textbf{Limitations} \\ [0.5ex] 
\hline 

\vfil Ref\cite{RL-SCOPF} & {\scriptsize \begin{itemize}
\itemsep0em
     \item \emph{States}: Grid Topology, Generator output, Loads and Line Flows
     \item \emph{Action Space}: Optimal topology for demand response, voltage modification based on voltage limits, generation rescheduling based on generation limits, load shedding
     \item \emph{Policy}: Optimizes rewards based on power balancing constraints, power flow equations, operational limits, and equipment capacity. 
 \end{itemize}} & {\scriptsize \begin{itemize}
 \itemsep0em
     \item Finds the level of vulnerability of a cyber-physical network to FDI attacks.
 \end{itemize}} & {\scriptsize \begin{itemize}
 \itemsep0em
        \item Does not provide a viable solution to protect against vulnerabilities to FDI attacks and assumes that the attacker has complete knowledge about the employed DRL and power system models to generate adversarial examples.
    \end{itemize}} \\ \hline

\vfil Ref\cite{SA1} & {\scriptsize\begin{itemize}
\itemsep0em
     \item \emph{States}: Network security situational awareness factors derived from power and information flows, and device states for both power and communication network. 
     \item \emph{Policy}: NN based optimization, with Gaussian distribution based output and activation functions for each layer 
 \end{itemize}}  & {\scriptsize\begin{itemize}
 \itemsep0em
        \item Able to detect and predict the effect of an attack on the communication network of a cyber-physical system. Utilizes the states of the elements in a connected network to extract situational awareness information.
    \end{itemize}}
          & {\scriptsize\begin{itemize} 
          \itemsep0em
        \item Does not account for the effect on the physical grid due to the cyber-attack on the communication network.
        \item Cannot withstand loss of communication from any agent.
    \end{itemize}}  \\ \hline

\vfil Refs\cite{marl_dos,ace} & {\scriptsize\begin{itemize}
\itemsep0em
     \item \emph{States}: System frequency and State of Charge (SoC) of the BESS in the network \cite{marl_dos}. Area Control Error (ACE) and load deviation \cite{ace}.
     \item \emph{Action Space}: Parameters of PI controllers and real power for BESS.  
 \end{itemize}}  & {\scriptsize\begin{itemize}
 \itemsep0em
        \item Able to withstand attacks to the communication network that blocks information flow between remote terminal units (RTUs) and BESS controllers by utilizing an event-triggered de-centralized secondary controller.
        \item Robust against communication topology changes in cases of communication dropout or an attack.
    \end{itemize}}
          & {\scriptsize\begin{itemize} 
          \itemsep0em
        \item The assumption that the DoS attack affects only the communication path between sensors and controllers may not hold in the real world. Also, variability in load has been ignored.
        \item Cannot withstand loss of communication from any agent.
    \end{itemize}}  \\ \hline

\vfil Ref\cite{RL-events,hmi} & {\scriptsize\begin{itemize}
\itemsep0em
     \item \emph{States}: Power generation, SoC of BESS, Loads \cite{RL-events}. Local load, local voltage magnitude, the energy storage state of electric vehicles (EVs) and thermostatically controlled loads (TCLs), and the utilization (whether EVs are connected) of EV \cite{hmi}.
     \item \emph{Action Space}: Binary power purchase decision from the grid, operation decision for PVs, charging/discharging of BESS, and load shedding \cite{RL-events}. Continuous charging/discharging power for EVs and TCLs constrained by their limits \cite{hmi}.
     \item \emph{Reward}: Robust reward function dependent on fuel prices, generation cost, operation cost and comfort level of loads. 
 \end{itemize}}  & {\scriptsize\begin{itemize}
 \itemsep0em
        \item Able to account for uncertainty in renewable generation, bad data from smart meters as well as extreme events. Adding robust optimization to DRL ensures a balance between optimality and feasibility of actions taken.
        \item Able to dramatically reduce the decision-making time by utilizing human knowledge and experience, and deal with events such as drop in generation, increase in load or topology changes.
    \end{itemize}}
          & {\scriptsize\begin{itemize} 
          \itemsep0em
        \item Does not account for effects of topology alterations (like line outages) inside the micro-grid.
        \item Provides a sub-optimal solution.
    \end{itemize}}  \\ \hline

\vfil Ref\cite{RL_Control1} & {\scriptsize\begin{itemize}
\itemsep0em
     \item \emph{States}: Voltage magnitudes and load levels at buses with controllable loads.
     \item \emph{Action Space}: Load shedding amount.
     \item \emph{Trainable Action Mask (TAM)}: Constrains action based on voltage stability criterion.
 \end{itemize}}  & {\scriptsize\begin{itemize}
 \itemsep0em
        \item Fast load shedding method is developed for rapid voltage stabilization in a network after an event occurs.
        \item Uses a guided evolutionary strategy and meta-learning for fast adaptability to changing operating conditions. 
    \end{itemize}}  & {\scriptsize\begin{itemize}
    \itemsep0em
        \item No controllability of the amount of load that needs to be shed (sheds a fixed 20\% of existing load).
        \item Does not consider control actions of other devices.
        \end{itemize}} \\ \hline
\end{tabular}   
\label{table:reinforcement}
\end{table*}

\section{Potential Challenges, Solutions and Path Forward} \label{disc}

The main reason for introducing physical principles in ML-based methods is to avoid infeasible solutions and address the lack of trustworthiness and interpretability {originating} from the black-box nature of ML algorithms. Although pure data-driven ML algorithms have achieved significant results, these methods cannot provide a high level of confidence for decision-makers to make appropriate decisions in dynamic systems. More specifically, control room operators need to evaluate the conditions of the system in real time and make decisions accordingly. The proposed PIML methods provide operators with an interpretation of the predictions and detection by ML algorithms. However, to address the main challenges in the dynamic systems and make the PIML methods more effective and applicable in real-world situations, the path forward needs to consider the following challenges {and the suggested solutions}:
\\

{\textbf{1) Spatio-temporal correlation extraction of measurement data:}}

{One of the major concerns of developing PIML methods for smart grid applications is their capabilities in capturing the spatial correlation of the system buses and the temporal evolution of dynamical systems. To address this issue, the following solutions are recommended:}

\begin{itemize}
\item{A combination of these methods can help address the spatial (location-dependent) and temporal (time-varying) correlations in the input data samples. As an example, the work in \cite{b41} employed a hybrid model of CNN and LSTM to extract spatial-temporal correlations for 4D trajectory prediction in aircraft. Since LSTM is very powerful in dealing with long-term dependencies, it is one of the commonly used algorithms for temporal feature extraction. The CNN algorithm was applied to capture the spatial correlations. In other instances, hybrid CNN models were used for powder-bed fusion process monitoring \cite{b42} and a combination of LSTM, CNN, and AE {was} employed for the detection of nontechnical power losses \cite{b43}.} 

\item{Another way of addressing spatial and temporal correlations by PIML methods is by using GNN as a type of physics-informed design of architecture NN with one of the variants of ML-based methods. Since GNNs are topology-aware, the topology of the system needs to be modeled through its elements to be suitable for the graph-based structure of GNNs as represented in \cite{b45} where elements of power distribution systems were modeled through nodes and edges of GNNs. These strategies are completely assessed and studied in \cite{b46}. Although there has been significant progress in PIML methods and their applications in power systems, if the proposed strategies like the ones developed in the above-mentioned papers are utilized for anomaly detection, classification, localization, and mitigation, they would not only help address issues of the black-box nature of ML methods but also appropriately deal with the spatiotemporal correlations in historical input data.}
\end{itemize}

{\textbf{2) System size and effectiveness of PIMLs:}}
It is worth mentioning that {since} the usage of some ML methods and their effectiveness depend on the system size \cite{b8}, different variants of PIML methods need to be applied to systems with different sizes to evaluate their effectiveness when the size of the systems changes.

{\textbf{3) Limited observability of systems:}}
One of the biggest limitations in the current state-of-the-art is the fact that most of them either assume complete observability of the system or do not deal with load variability in the network. However, for continuous operation in the real world, lack of observability (due to lack of sensors, lapse in communication, or topology alteration) and variability (which is an inherent aspect of all loads) have to be taken into account. The path forward should not only take into account the spatiotemporal aspect of data obtained from sensors but utilize physical knowledge of power networks to address the complex nature of {these} critical systems that {are} the backbone of today's society.

\section{Summary} \label{concl}
In this work, {the applications of} PIML methods for anomaly detection, classification, localization, and mitigation in power transmission and distribution systems were thoroughly reviewed. The strategies for integrating physical {laws} into different types of ML methods, i.e., supervised/semi-supervised, unsupervised, and RL, were assessed in detail with relevant {references from different scientific fields to clarify their implementation approaches for the interested readers}. To make the {existing} PIML methods more practical in real-world {applications, we discussed the potential challenges, their current capabilities, and possible solutions that shed light on the future research directions in the more reliable and secure operation of cyber-power systems.}

\bibliographystyle{ieeetr}
\bibliography{References.bib}

\begin{IEEEbiography}[{\includegraphics[width=1.05in,height=1.25in,clip,keepaspectratio]{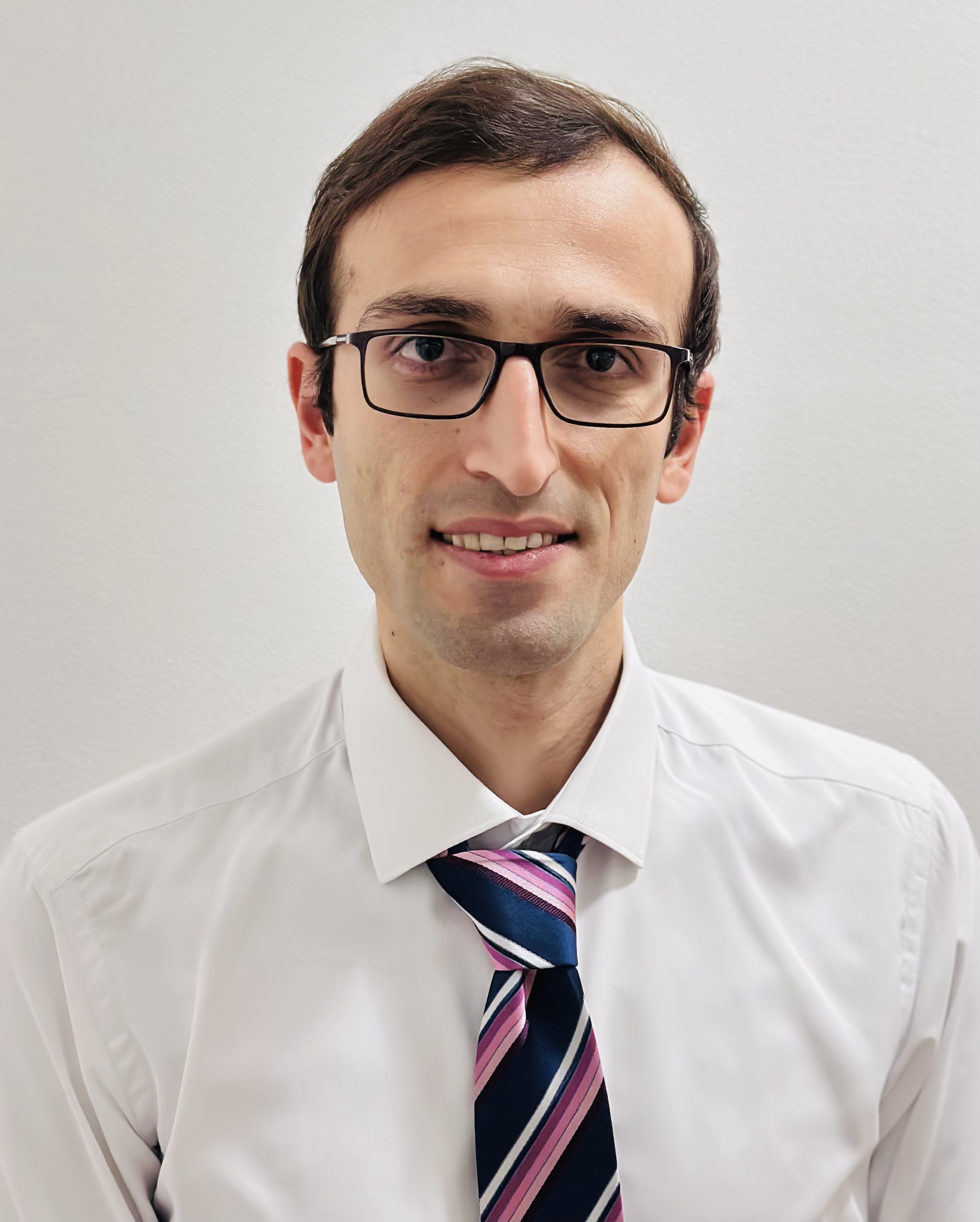}}]{Mehdi Jabbari Zideh} (Student Member, IEEE) received his B.Sc. degree from the University of Tabriz, Tabriz, Iran in 2014 and his M.Sc. degree from the University of Guilan, Rasht, Iran, in 2017, both in electrical engineering. Since 2022, he has been working toward his Ph.D. degree with the Lane Department of Computer Science and Electrical Engineering, West Virginia University, Morgantown, WV, USA. His research interests include applications of machine learning in power system operation, monitoring, and optimization, cyber-physical systems analysis, distribution systems, and smart grids.
\end{IEEEbiography}

\begin{IEEEbiography}[{\includegraphics[width=1.05in,height=1.25in,clip,keepaspectratio]{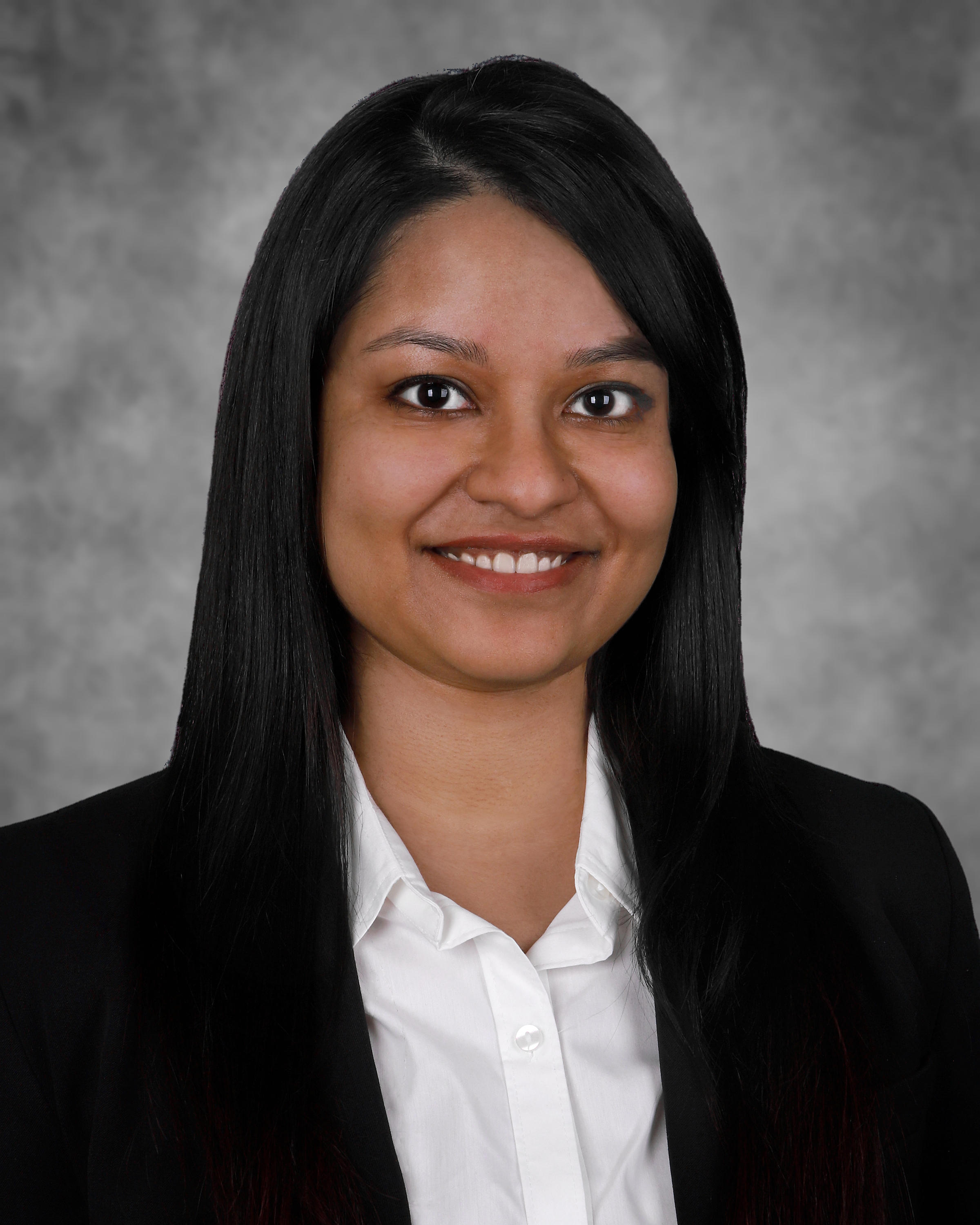}}]{Paroma Chatterjee}(Member, IEEE) received her Bachelors from Jadavpur University, Kolkata, India in 2012 and Masters in Electrical Engineering from Virginia Tech, Blacksburg, USA, and Arizona State University, Tempe, USA in 2015 and 2020, respectively. She has worked as a Transmission Analyst, providing services to banks/equity investors, developers, and utilities, primarily focused on impacts of electric high voltage transmission on clients’ financial, development, reliability, planning, and economic decisions related to generation and transmission assets, with Transmission Analytics Consulting, LLC, Phoenix, USA. She is currently working toward her Ph.D. in Electrical Engineering at West Virginia University, Morgantown, WV, USA. Her research interests include machine-learning based power system monitoring and control for improved grid resiliency.
\end{IEEEbiography}

\begin{IEEEbiography}[{\includegraphics[width=1.05in,height=1.25in,clip,keepaspectratio]{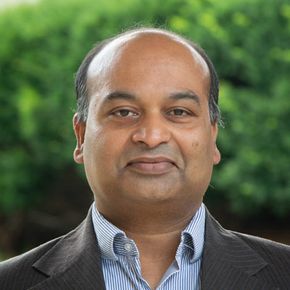}}]{Anurag K. Srivastava} (Fellow, IEEE) received the Ph.D. degree in electrical engineering from the Illinois Institute of Technology, Chicago, IL, USA, in 2005. He is the Raymond J. Lane Professor and Chairperson of the Computer Science and Electrical Engineering Department, West Virginia University, Morgantown, WV, USA. He is also an Adjunct Professor with the Washington State University, Pullman, WA, USA, and Senior Scientist with the Pacific Northwest National Lab, Richland, WA. His research interests include data-driven algorithms for power system operation and control, including resiliency analysis. He is the Vice-Chair of IEEE PES power system operation SC, and Co-Chair of tools for power grid resilience TF. He is the PES Distinguished Lecturer and the author of more than 300 technical publications, including a book on power system security and three patents.
\end{IEEEbiography}


\end{document}